\documentclass[a4paper,11pt]{article}
\pdfoutput=1 % if your are submitting a pdflatex (i.e. if you have
\usepackage{jheppub} % for details on the use of the package, please
\usepackage{amsmath}
\usepackage{amssymb}
\usepackage{dsfont}	
\usepackage{braket}
\usepackage{tensor}
\allowdisplaybreaks
\usepackage[mathscr]{euscript}
\usepackage{tikz}
\usetikzlibrary{decorations.pathmorphing}
\usetikzlibrary{hobby}

\tikzset{snake it/.style={decorate, decoration=snake}}

\numberwithin{equation}{section}

\def\be{\begin{equation}}
\def\ee{\end{equation}}

\newcommand{\nif}{m} % number of insertion fields

\title{Correlators of the symmetric product orbifold}

\author[a]{Andrea Dei}
\author[b]{\!\!\!, Lorenz Eberhardt}
\affiliation[a]{Institut f\"ur Theoretische Physik, ETH Zurich, \\
\hspace*{0.3cm}CH-8093 Z\"urich, Switzerland}
\affiliation[b]{School of Natural Sciences, Institute for Advanced Study, \\
\hspace*{0.3cm}Princeton, NJ 08540, USA}
\emailAdd{adei@itp.phys.ethz.ch}
\emailAdd{elorenz@ias.edu}

\abstract{We exploit null vectors of the fractional Virasoro algebra of the symmetric product orbifold to compute correlation functions of twist fields in the large $N$ limit. 
This yields a new method to derive correlation functions in these orbifold CFTs that is purely based on the symmetry algebra. 
We explore various generalisations, such as subleading (torus) contributions or correlation functions of other fields than the bare twist fields.
We comment on the consequences of our computation for the $\text{AdS}_3/\text{CFT}_2$ correspondence.
}

\begin{document}
\maketitle
\flushbottom

\section{Introduction}
The symmetric product orbifold CFT has played a major role in the $\text{AdS}_3/\text{CFT}_2$ correspondence and served as a laboratory to explore fundamental aspects of holography, see e.g.~\cite{Dijkgraaf:1996xw, Maldacena:1997re, Seiberg:1999xz, Dijkgraaf:2000fq, Hartman:2014oaa, Gaberdiel:2014cha, Gaberdiel:2015mra} for an incomplete list, and \cite{Aharony:1999ti, David:2002wn} for reviews. Numerous checks have provided evidence for the fact that the dual CFT to string theory on $\text{AdS}_3\times \mathrm{S}^3 \times \mathcal{M}_4$ for $\mathcal{M}_4=\mathbb{T}^4$ or $\mathrm{K3}$ is on the same conformal manifold as the symmetric product orbifold $\text{Sym}^N(\mathcal{M}_4)$ \cite{Dijkgraaf:1998gf,deBoer:1998us,Maldacena:1999bp,Larsen:1999uk,Seiberg:1999xz,Argurio:2000tb,Gaberdiel:2007vu,Dabholkar:2007ey}.  

A more refined incarnation of the duality has recently been formulated, where the precise string background dual to the symmetric orbifold (and not just its moduli space) has been identified.
The proposal is most stringent for string theory on $\text{AdS}_3 \times \text{S}^3 \times \mathbb{T}^4$ with one unit of NS-NS flux, where the dual CFT is given by the symmetric product orbifold of the sigma-model on $\mathbb{T}^4$. In this instance, the full perturbative string spectrum was matched on both sides \cite{Gaberdiel:2018rqv, Eberhardt:2018ouy}. Moreover, the structure of correlation functions was discussed in \cite{Eberhardt:2019ywk} and it was found that the worldsheet correlation functions have the correct form to reproduce the symmetric product orbifold ones. 

In \cite{Eberhardt:2019qcl}, a natural extension of this idea was proposed. It was noticed that for higher values of NS-NS flux, the perturbative long-string spectrum matches that of a symmetric product orbifold of $\mathcal{N}=4$ Liouville theory tensored with $\mathbb{T}^4$. By the same token, the perturbative long-string sector of bosonic string theory on $\text{AdS}_3$ matches with a symmetric product orbifold of bosonic Liouville theory tensored with a suitable internal CFT. This proposal was further tested in \cite{Dei:2019osr}, where null-vector constraints on correlation functions were matched. This implies the matching of those correlation functions that are fixed solely by the existence of these null vectors, up to a normalisation constant.
\medskip

Two-dimensional conformal field theories possess an infinite-dimensional symmetry algebra. 
Correlation functions in 2d CFTs can sometimes be computed exactly by exploiting the constraints imposed on them by the respective symmetry algebra. Such constraints arise in particular from the existence of null vectors. 
For instance, it is well-known that for Virasoro minimal models, Ward identities 
lead to a differential equation for the four-point function of primary fields \cite{Belavin:1984vu}. Analogous constructions exist for different symmetry algebras, for example the Knizhnik-Zamolodchikov equations for affine symmetry  \cite{Knizhnik:1984nr}, or for $\mathcal{W}_3$ \cite{Zamolodchikov:1985wn,Fateev:1987vh,Bowcock:1992gt,Gaberdiel:1993mt, Fateev:2007ab}. For $\mathcal{W}_3$, this becomes significantly harder and one needs two independent null vectors. For generic $\mathcal{W}$-algebras, one needs several independent null vectors to turn the constraints into differential equations. 
There seems to be no general recipe whether and how null-vector constraints can be turned into differential equations for arbitrary symmetry algebras.

In this paper, we study orbifold CFTs and specifically the symmetric product orbifold. The symmetric product orbifold has a very large symmetry algebra -- the so-called ``Higher Spin Square'' \cite{Gaberdiel:2015mra, Gaberdiel:2015wpo}.\footnote{More precisely, the higher spin square is the symmetry algebra of the symmetric product orbifold of a single free boson.} Twisted sectors of the orbifold carry a twisted representation of this algebra, whose moding is fractional. One might suspect that the situation is similar to the one of $\mathcal{W}$-algebras, and that the symmetry algebra alone is not enough to fully constrain correlation functions. For the symmetric orbifold, we will show that contrary to this expectation, one can sometimes turn null-vector constraints into differential equations for correlators. We believe that our reasoning can be extended to other permutation orbifolds.

Correlators in the symmetric orbifold were first considered and computed in \cite{Arutyunov:1997gt, Arutyunov:1997gi} by using the stress-tensor method of \cite{Dixon:1986qv}. This was then applied to compute correlators of chiral primaries in the symmetric orbifold of $\mathbb{T}^4$ in \cite{Jevicki:1998bm}.
The idea of the stress-tensor method is to determine the correlator of twist fields with one stress-energy tensor insertion by employing a free field construction (in terms of free bosons or fermions). Subsequently, the correlator is fixed by imposing all monodromy properties. This in turn leads to a differential equation for the original correlator of twist fields, which is hence fixed up to an overall constant.

For the symmetric product orbifold, a powerful alternative method to compute the correlators was developed by Lunin and Mathur \cite{Lunin:2000yv, Lunin:2001pw}. It relies on the fact that while correlators of twist fields introduce monodromies for the fundamental fields in the orbifold, these monodromies may be trivialised by lifting the problem to a suitable covering surface. The computation of the correlation functions of twist fields can thus be related to a correlation function on the (potentially higher genus) covering surface. 

While both methods have their strengths, we find it highly desirable to derive the same result directly from the symmetry algebra and without invoking free field realisations, complicated conformal transformations, regularisation of the Liouville action, etc. Our main motivation for this is holography; since the null-vector constraints were matched in \cite{Dei:2019osr}, such a derivation would effectively demonstrate the matching of correlation functions (at genus 0) in this instance of the $\text{AdS}_3/\text{CFT}_2$ correspondence.
\medskip

We will develop such a method to compute correlation functions in the symmetric product orbifold theory that makes only use of the symmetry algebra and does not require any free field realisation or regularisation. With this, we reproduce and extend the results about correlators found in the literature~\cite{Arutyunov:1997gt, Lunin:2000yv, Pakman:2009zz}. We should also note that our derivation of the correlation functions is quite short and elegant. On the flip side, since we are fixing the correlators via differential equations, there are normalisation factors that cannot be computed from null vectors alone. We determine them by requiring factorisation of the four-point function into three-point functions.

The basic idea is the following. The symmetric orbifold has a fractionally moded Virasoro algebra $L_m$ with $m \in \tfrac{1}{w}\mathds{Z}$ acting on its twisted sector (which is parametrised by a single cycle conjugacy class $[(1 \cdots w)]$). Analogously to the obvious null vector $L_{-1}\ket{0}$ in the untwisted sector, the theory has the null vector $L_{-1/w} \ket{\sigma_w}$ in the twisted sector, where $\ket{\sigma_w}$ is the twisted sector ground state. Thus, when computing, say, a correlation function of four twist fields, we have the constraint
\be 
\big\langle (L_{-\frac{1}{w_1}}\sigma_{w_1})(z_1) \sigma_{w_2}(z_2) \sigma_{w_3}(z_3) \sigma_{w_4}(z_4) \big\rangle=0\ . 
\ee
It is now however not obvious how one would turn this into a differential equation for the correlator of the four twist fields. On the face of it, there are many terms in the OPE of this fractional Virasoro algebra with the twist fields that we cannot evaluate. We show however that there is a very simple criterion that makes this condition evaluable. Namely, when the corresponding covering space is a sphere, one can turn this constraint into a differential equation. We should note that similar methods were already applied in the context of $\mathds{Z}_2$ orbifolds, see e.g. \cite{Friedan:1984rv, Gaberdiel:1996kf, Dupic:2017hpb}.

From a holographic point of view, we are hence able to compute correlators in the dual CFT (which is defined on the sphere), whose covering space is also a sphere. In the AdS/CFT correspondence, the covering space becomes identified with the string worldsheet \cite{Pakman:2009zz, Eberhardt:2019ywk} and hence these correlators correspond to the leading (connected) terms in the large $N$ expansion. This should be expected, since these correlators are exactly those that can be computed on the worldsheet by BPZ-like differential equations. 

We also demonstrate that the method is applicable beyond the twisted sector ground states. Again inspired by holography, we derive the analogue of the BPZ equation in the twisted sector for a degenerate field which has a null vector at level 2. In all instances we compute, we show that the structure of the result is always controlled by the covering map as follows. For fields in the respective twisted sector, we find that the solution to the differential equations can be written as
\be
\left\langle \prod_{j=1}^m\mathcal{O}_{w_j}(z_j) \right\rangle 
= \text{covering map data} \times \left\langle \prod_{j=1}^m\tilde{\mathcal{O}}(t_j) \right\rangle\ , \label{eq:relation base space covering space}
\ee
i.e.~the correlator in the orbifold CFT can be expressed in terms of the correlator in the covering space (which by assumption is also a sphere and which we have parametrised by the coordinate $t$). The precise form of this formula is given in eq.~\eqref{eq:precise relation base space covering space}. The factor, which we schematically denoted by `covering map data' can be interpreted as the conformal factors from the conformal transformation to the covering space in the covering space method of Lunin and Mathur. We stress however that we get this result purely from the symmetry algebra.

We also show that one can perform the same analysis when the covering space is a torus, but to evaluate the constraint one needs a second null vector. This requires the theory of which we are taking the symmetric orbifold to be rational and, slightly extending the results of \cite{Dupic:2017hpb}, we consider Virasoro minimal models for illustration. We analyse the perhaps simplest case, where the original CFT is the Ising model in detail. In particular, we review how to derive a differential equation for the correlator
\be 
\langle \sigma_2(z_1) \sigma_2(z_2) \sigma_2(z_3) \sigma_2(z_4)\rangle
\ee
in the theory $(\text{Ising}\times\text{Ising})/\mathds{Z}_2$, whose covering space is a torus. By the covering space method, this correlator is related to the torus partition function of the Ising CFT. This relation is well-known and was used for instance recently in \cite{Hartman:2019pcd}. Thus, this gives an easy method to derive modular differential equations for rational CFTs; the corresponding modular differential equations are reinterpreted as ordinary differential equations for the four-point function of twist fields.
\medskip

This paper is organised as follows. In Section~\ref{sec:symm orbifold}, we start by reviewing some important features of (non-abelian) orbifolds and consider in detail the special case of the symmetric orbifold. We also explain the large $N$ expansion in the symmetric orbifold and the relation to string theory on $\text{AdS}_3$. The heart of our argument is spelled out in Section~\ref{sec:diffeq-ssss}, where we derive a differential equation for the four-point function of twisted sector ground states. We solve the equation in full generality analytically, which fixes the four-point function up to an overall constant. We then show in Section~\ref{sec:fixing integration constant} how to fix this integration constant by imposing factorisation of the four-point function into different OPE channels. In the process we explore a very convenient parametrisation of the covering map in terms of its pole locations \cite{Roumpedakis:2018tdb}.
Since the discussion of Sections~\ref{sec:diffeq-ssss} and~\ref{sec:fixing integration constant} is a bit abstract, we then exemplify in Section~\ref{sec:example} the construction with the example of the four-point function where all the fields have twist two. In the following two sections, we explore generalisations of the basic argument. In Section~\ref{sec:more-differential-equations}, we consider correlation functions, where the fields are not necessarily twisted sector ground states. We consider the case where one of the fields has a null vector at level 2. We also generalise the analysis to higher-point functions. Finally, in Section~\ref{sec:higher genus} we discuss how far the method can be extended to the case where the covering space has higher genus. We end with a discussion and future directions in Section~\ref{sec:discussion}. Section~\ref{sec:discussion} contains also a discussion about the implications of our result for the worldsheet theory dual to the symmetric product orbifold.

\section{The symmetric product orbifold}\label{sec:symm orbifold}

Let us set up our notation by recalling some facts about the conformal field theory of orbifolds, the symmetric product orbifold and its relation to string theory on $\text{AdS}_3$. Readers that only want to see our argument should jump directly to Section~\ref{sec:diffeq-ssss}.

\subsection{Conformal field theory of orbifolds}

Consider a conformal field theory sigma-model whose target space is the manifold $\mathscr{M}$ with a finite symmetry group $G$.\footnote{The following discussion also goes through if the CFT is not a sigma-model or when $G$ is not a finite group.} One can define a new conformal field theory on the orbifold 
\be 
\mathscr{M}\big/ G
\ee 
by identifying points related by the action of $G$, i.e. 
\be 
x \sim g \cdot x \ ,\qquad x \in \mathscr{M} \ , \qquad g \in G \ ,
\label{eq:x=gx}
\ee
which corresponds to a gauging of the discrete symmetry.
The CFT on $\mathscr{M}/G$ contains fields obeying twisted boundary conditions, 
\be 
X(e^{2 \pi i} z) = (g\cdot X)(z) \ , 
\label{eq:twisted-bondary-cond}
\ee
where here and in the following $\mathrm{e}^{2\pi i}z$ stands as a short hand for the monodromy of $z$ around 0. 
Consider $g, h \in G$. Since $X$ and $h \cdot X$ are in the same gauge orbit and $h \cdot X$ satisfies
\be 
(h \cdot X)(e^{2 \pi i} z) = (hgh^{-1}\cdot (h \cdot X))(z) \ , 
\ee 
twisted sectors are labeled by the conjugacy classes of $G$. We will denote in the following the conjugacy class of the group element $g$ by $[g]$.

Equation~\eqref{eq:twisted-bondary-cond} is implemented by introducing twist fields $\sigma_g(z)$, so that fields acquire monodromies, 
\be 
X(e^{2 \pi i}z + \zeta) \sigma_g(\zeta) = (g\cdot X)(z+\zeta) \sigma_g(\zeta) \ , \qquad g \in G \ . 
\label{eq:twisted-bound-cond}
\ee
Here we put the twist field at a generic location $\zeta$. 

\begin{figure}
\begin{minipage}{1\textwidth}
\centering
\makebox[0pt]{%
\begin{tikzpicture}

\useasboundingbox (-3.5,-3.5) rectangle (11.5,2.5);
%\draw (-3.5,-3.5) rectangle (11.5,2.5);
\draw[thick, red,bend left=20, snake it] (-1,-.3) to (-2,2);
\draw[thick, red,bend right=20, snake it] (1,-.3) to (2,2);
\fill (-1,-0.3) circle (.1) node[below] {$\sigma_{g_2}(z_2)$};
\fill (1,-0.3) circle (.1) node[below] {$\sigma_{g_1}(z_1)$};
\draw[thick, ->]  (2.3,-2.4)  to [curve  through ={(2,0.4) . . (1.3,0.6) . . (0.1,-0.8)  }]  (0.2,-1.4) ;
\draw[thick, <-]  (-2.3,-2.4)  to [curve  through ={(-2,0.4) . . (-1.3,0.6) . . (-0.1,-0.8)  }]  (-0.2,-1.4) ;
\draw[thick,-> ] (2.5,-2.5) to [curve through={(3,-2). . (0,1.5) . . (-3,-2)}] (-2.5,-2.5);
\node at (0,-1.8) {$X^{(g_1(j))}(z)$};
\node at (2,-2.8) {$X^{(j)}(z)$};
\node at (-2,-2.8) {$X^{(g_2g_1(j))}(z)$};

\draw[thick, red,bend left=20, snake it] (7,-.3) to (6,2);
\draw[thick, red,bend right=20, snake it] (9,-.3) to (10,2);
\fill (7,-0.3) circle (.1) node[below] {$\sigma_{g_1'}(z_1)$};
\fill (9,-0.3) circle (.1) node[below] {$\sigma_{g_2'}(z_2)$};

\draw[thick,-> ] (10.5,-2.5) to [curve through={(11,-2). . (8,1.5) . . (5,-2)}] (5.5,-2.5);
\draw[thick,-> ] (8.1,-1.4) to [curve through={(9,-1.1) . . (9.5,-1.2) . . (9.6,0) . . (8.5,0) . . (8,-1) . . (6,-1.7)}] (5.8,-2.4);
\draw[thick,-> ] (10,-2.5) to [curve through={(11,-1) . . (10,0.7) . . (8,1.2) . . (6,-0.3) . . (7.5,-1) . . (8,-0.1)  .  . (9.5,0.4) . . (9.5,-1.5) . . (9,-1.3)}] (8.3,-1.5);
\node at (8,-1.8) {$X^{(g_2'^{-1}g_1'g_2'(j))}(z)$};
\node at (10,-2.8) {$X^{(j)}(z)$};
\node at (6,-2.8) {$X^{(g_1'g_2'(j))}(z)$};
\end{tikzpicture}}
\end{minipage}
\caption{Monodromy of the twist fields. We drew our choice of branch cuts as red wiggly lines. Every time the contour crosses the branch cut (the symmetry defect), the field picks up the mondromy of the twist field.} \label{fig:monodromy twist fields}
\end{figure}
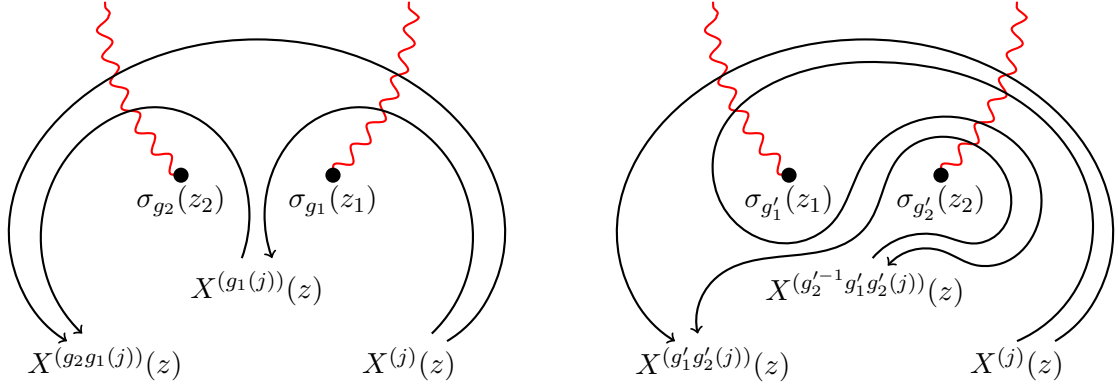

The twist fields have themselves monodromy around each other. To see this, let us insert two twist fields in the plane and let us consider the mondromy of the field $X(z)$ around them. We then rotate the twist fields by $180^\circ$, which drags the contours around as shown in Figure~\ref{fig:monodromy twist fields}. After half a rotation, the group elements associated to the twist fields can be read off from the figure and are
\be 
g_1'=g_2 g_1 g_2^{-1}\ , \qquad g_2'=g_2\ . \label{eq:g primes}
\ee
After another $180^\circ$, the process is iterated and we find that
\be 
\sigma_{g_2}(e^{2 \pi i}z + \zeta) \sigma_{g_1}(\zeta)=\sigma_{g g_2 g^{-1}}(z + \zeta) \sigma_{g g_1 g^{-1}}(\zeta)\ ,
\label{eq:sigma-sigma-monodromy}
\ee
where $g=g_2 g_1$. 

We should stress that the fields $\sigma_g(z)$ are not local fields, since their correlation functions will involve branch cuts. As a consequence, their OPEs are not symmetric. This phenomenon is well-known, for an extensive review see e.g.~\cite{Bhardwaj:2017xup} and references therein.

Hence the twist fields $\sigma_g(z)$ are themselves not gauge invariant, i.e.~they depend on the specific element $g \in G$. Gauge invariant twist operators can be obtained by summing over the orbit of the group, 
\be 
\sigma_{[g]}(z)\equiv \mathcal{N}_{[g]} \sum_{h \in G} \sigma_{hgh^{-1}}(z) \label{eq:gauge invariant sigma} \ , 
\ee
where
\be 
\mathcal{N}_{[g]}=\frac{1}{\sqrt{|\text{Stab}([g])| |G|}}
\label{eq:normalization-factor}
\ee
is a normalisation factor and $[g]$ denotes the conjugacy class with representative $g$. The normalisation factor \eqref{eq:normalization-factor} ensures that $\sigma_{[g]}(z)$ has a properly normalised two-function (assuming that $\sigma_g(z)$ has a normalised two-point function). 

It should be clear from eq.~\eqref{eq:twisted-bound-cond} and Figure~\ref{fig:monodromy twist fields} that for $z_1 \sim z_2$, $\sigma_{g_1}(z_1) \sigma_{g_2}(z_2)$ and $\sigma_{g_1 g_2}(z_2)$ implement the same boundary conditions on fields. In other words, states in the conjugacy classes of $g_1$ and $g_2$ fuse to states in the conjugacy class of $g_1 g_2$ \cite{Hamidi:1986vh,Dixon:1986qv,Dijkgraaf:1989hb, Arutyunov:1997gt} and we have the OPE 
\be 
\sigma_{g_1}(z_1) \sigma_{g_2}(z_2) \sim \tensor{C}{_{g_1, g_2}^{g_1g_2}} |z_1-z_2|^{-2(h_{g_1} + h_{g_2} - h_{g_1g_2})} \sigma_{g_1 g_2}(z_2) + \dots 
\label{eq:sigma-sigma-OPE}
\ee     
where $\tensor{C}{_{g_1, g_2}^{g_1g_2}}$ are the structure constants and the dots denote higher orders in $(z_1-z_2)$. $h_{g_1}, h_{g_2}, h_{g_1g_2}$ are the conformal weights of $\sigma_{g_1}$, $\sigma_{g_2}$, $\sigma_{g_1 g_2}$ respectively.  Note that this OPE is exactly compatible with the exchange of the two twist fields we derived above. Upon exchanging the two fields, $g_1 \to g_1'$ and $g_2\to g_2'$, see eq.~\eqref{eq:g primes}. Since $g_1'g_2'=g_2g_1$, this is compatible with the RHS of the OPE. In the following we shall always fix one order in which we take the OPE, or equivalently one choice of branch cuts, as in Figure~\ref{fig:monodromy twist fields}. One convenient way of fixing the ordering is radial ordering, i.e.~$|z_1|<|z_2|$, see also \cite{Pakman:2009zz} for a discussion on this.

Consider three conjugacy classes $[g_1]$, $[g_2]$ and  $[g_3]$. Then the fusion rules can be written as
\be 
[g_1] \times [g_2] = \sum_{[g_3]} \tensor{N}{_{[g_1][g_2][g_3^{-1}]}} [g_3] \ ,
\label{eq:fusion-rules-abstract}
\ee 
where the sum runs over all conjugacy classes. We used a symmetric notation for the fusion coefficient and made use of the fact that $\sigma_{g^{-1}}(z)$ is the conjugate field to $\sigma_g(z)$.
The fusion coefficient takes the form \cite{Dijkgraaf:1989hb}
\be 
\tensor{N}{_{[g_1][g_2][g_3]}}= \bigg|\{(g_1,g_2,g_3)\, |\, g_1g_2g_3=\mathds{1} \}/\sim \bigg|\ ,\label{eq:fusion coefficient}
\ee
where
\be 
(g_1,g_2,g_3) \sim (g g_1 g^{-1}, g g_2 g^{-1}, g g_3 g^{-1})
\ee
for any group element $g$.

One can study correlation functions of gauge invariant operators  
\be 
\left \langle \sigma_{[g_1]}(z_1) \dots \sigma_{[g_\nif]}(z_\nif) \right\rangle \ , 
\label{eq:gauge-inv-corr}
\ee 
by exploiting \eqref{eq:gauge invariant sigma} and decomposing \eqref{eq:gauge-inv-corr} into a sum of correlation functions of gauge dependent twist fields of the form 
\be 
\left \langle \sigma_{g_1}(z_1) \dots \sigma_{g_\nif}(z_\nif) \right\rangle \ ,
\label{eq:gauge-dep-corr}
\ee
i.e., 
\be 
\left \langle \sigma_{[g_1]}(z_1) \dots \sigma_{[g_\nif]}(z_\nif) \right\rangle = \prod_{j=1}^\nif \mathcal{N}_{[g_j]}  \sum_{h_1, \dots, h_\nif \in G} \left \langle \sigma_{h_1g_1h_1^{-1}}(z_1) \dots \sigma_{h_\nif g_\nif h_\nif^{-1}}(z_\nif) \right\rangle \ ,
\label{eq:gauge-inv=sum-gauge-dep}
\ee
These will be analysed in detail for the symmetric orbifold in Section \ref{subsec:large N}. 

Equation \eqref{eq:sigma-sigma-monodromy} shows explicitly that correlation functions of gauge dependent twist operators are not single-valued. In fact, only the orbifold projection (imposed by the sum in \eqref{eq:gauge-inv=sum-gauge-dep}) ensures single-valuedness. Indeed, since the monodromy of twist fields relates elements inside the same conjugacy class (see eq.~\eqref{eq:sigma-sigma-monodromy}), correlation functions of gauge invariant twist fields are single-valued. When performing a monodromy transformation, different gauge dependent contributions entering the sum \eqref{eq:gauge-inv=sum-gauge-dep} get reshuffled. 

\subsection[The symmetric orbifold and string theory on \texorpdfstring{AdS$_3$}{AdS3}]{The symmetric orbifold and string theory on AdS$_{\boldsymbol{3}}$}\label{subsec:AdS3}

Let us proceed by specialising our analysis to the symmetric product orbifold
\be
\text{Sym}^N(\mathcal{M}) \equiv \left.\mathcal{M}^{\otimes N} \right/ S_N\ .
\ee
In the language of the previous section, we have $\mathscr{M} = \mathcal{M}^{\otimes N}$, i.e.~we quotient the CFT living on $\mathcal{M}^{\otimes N}$ by the action of the symmetric group $S_N$ on the different $N$ copies. We will refer to the CFT on $\mathcal{M}$ as the \emph{seed theory}. It has central charge $c$, so that the full symmetric product orbifold has central charge $Nc$. The $N$ copies of the seed theory are permuted under the action of the symmetric group $S_N$, 
\be 
g:\, X^{(k)} \longrightarrow X^{(g(k))} \ , \qquad g \in S_N \ , 
\label{eq:twisted-bound-cond-symm}
\ee
where $X^{(k)}$ is a generic field in the $k$-th copy of the CFT on $\mathcal{M}$. Twisted sectors are in one to one correspondence with conjugacy classes of $S_N$, which in turn are characterised by their cycle shapes. 
\smallskip

In the following we will be interested in holographic applications of the symmetric orbifold CFT. This will require us to concentrate on the large $N$ limit of the orbifold, which describes perturbative string theory on $\mathrm{AdS}_3$ backgrounds \cite{Maldacena:1997re, Eberhardt:2018ouy}. The large $N$ limit of the symmetric orbifold will be reviewed in Section \ref{subsec:large N}. Moreover, again in view of holographic applications, we will mostly restrict to twisted sectors described by conjugacy classes of single cycles. A multi-cycle correlator can be obtained as a coincidence limit of a single cycle correlator, but possibly with more twist fields inserted. We will make further comments on this in Section \ref{sec:higher-point-and-multi-cycle-corr}. Let us explain how this comes about. The cycle structure of $S_N$ and the Hilbert space of the symmetric orbifold have the structure of a Fock space of identical particles. This reflects the fact that this CFT is conjectured to describe string (field) theory on $\mathrm{AdS}_3$ and under this correspondence the single particle states are identified with single string states on $\mathrm{AdS}_3$. In particular, single cycle twist fields are interpreted as single string states. Thus, in the following, we will describe the twisted sector simply by an integer $w \in \mathds{Z}_{\ge 1}$, corresponding to the length of the single cycle
\be 
w \equiv [(1 \cdots w)]\ ,
\ee
and by the same token $\sigma_w\equiv \sigma_{[(1 \cdots w)]}$.
Twist fields of single cycle shape have conformal weight 
\be 
h_w^0 = \bar h_w^0 = \dfrac{c(w^2-1)}{24w} \ , 
\label{eq:conformal-weight-single-cycle}
\ee
where $c$ is the central charge of the seed theory on $\mathcal{M}$. We will refer to $w$ as the `twist' of the twist field. 

Consider a single cycle permutation $g$ of length $w$. The energy-momentum tensors are just a special case of \eqref{eq:twisted-bound-cond-symm} and hence in the presence of a twist field, the different copies are related by 
\be 
T^{(k)}(\mathrm{e}^{2 \pi i}z + \zeta) \sigma_g(\zeta)= T^{(g(k))}(z + \zeta) \sigma_g(\zeta)\ .
\ee 
One can diagonalise the action of the twist field by defining
\be 
T^{\ell}(z) \equiv \sum_{r=0}^{w-1} \mathrm{exp}\left(-\frac{2 \pi i \ n_r \ \ell }{w}\right) T^{(n_r)}(z) \ , \qquad \ell= 0 , \dots , w-1\ , \label{eq:Fourier transformed Ts}
\ee
where $g= (n_1 \cdots n_w)$.\footnote{Eq.~\eqref{eq:Fourier transformed Ts} is only well-defined up to a phase. This reflects the freedom to cyclically permute the elements in the cycle $g$. We fix this phase conventionally by choosing $n_1$ to be the smallest element in the cycle.} In other words, we sum over the orbit weighted with phases. These Fourier-transformed operators satisfy
\be 
T^{\ell}(e^{2 \pi i}z + \zeta)\sigma_g(\zeta) = \mathrm{e}^{\frac{2 \pi i \ell}{w}} T^\ell(z + \zeta) \sigma_g(\zeta)\ ,\label{eq:monodromy T}
\ee
and hence transform diagonally under the twist. For concreteness, let us focus on the case $g=(1\cdots w)$, but it should be clear that all the definitions exist for arbitrary cycles, since they can be obtained by a relabelling of $1$, $\dots$, $w$.
We define fractional Virasoro modes by\footnote{Note that the integrand is single-valued in the presence of a twist-field thanks to the monodromy \eqref{eq:monodromy T}.}
\be 
L_n^{(1\cdots w)} = \oint \mathrm{d}z \  z^{n+1}\ T^\ell(z) \ , \qquad n \in \mathbb{Z} - \tfrac{\ell}{w} \ . 
\label{eq:fractional Virasoro modes}
\ee
The OPE of the fields $T^\ell$ with ground state twisted sectors is 
\be  
T^\ell(z) \sigma_{g}(u) =  \dfrac{L_{-\ell/w}^{(1\cdots w)}\sigma_{g}(u)}{(z-u)^{2-\frac{\ell}{w}}} + \dfrac{L_{-1-\ell/w}^{(1\cdots w)}\sigma_{g}(u)}{(z-u)^{1-\frac{\ell}{w}}} + \dots
\label{eq:OPE-Tsigma}
\ee 
The fractional Virasoro generators \eqref{eq:fractional Virasoro modes} still satisfy the Virasoro algebra:
\be 
[L^{(1\cdots w)}_m,L^{(1\cdots w)}_n]=(m-n)L^{(1\cdots w)}_{m+n}+\frac{c\, w}{12}m(m^2-1)\delta_{m+n,0}\ .
\ee
This algebra is in fact isomorphic to the Virasoro algebra of a single copy, as can be seen by defining
\be 
L^{(1\cdots w)}_m=\frac{1}{w} \hat{L}_{m w}+\frac{c(w^2-1)}{24w}\delta_{m,0}\ ,
\label{eq:Virasoro-algebra-isomorphism}
\ee
where the generators $\hat{L}_m$ are the Virasoro generators of a single copy of $\mathcal{M}$ and have central charge $c$.
This is the expected transformation behaviour associated to the conformal transformation $z \to z^w$, as was worked out explicitly in \cite{Burrington:2018upk}. We often suppress the label $(1\cdots w)$ for the Virasoro generator below, expressions like $L_{-1/w} \sigma_w$ are understood to mean $L^{(1 \cdots w)}_{-1/w} \sigma_w$.

\subsection[Large \texorpdfstring{$N$}{N} expansion]{Large $\boldsymbol{N}$ expansion} \label{subsec:large N}
Correlation functions in the symmetric product orbifold have a large $N$ expansion, which we shall now review \cite{Lunin:2000yv, Pakman:2009zz}. It essentially originates from the normalisation in \eqref{eq:gauge invariant sigma}, that for single cycles of length $w$ amounts to 
\be 
\sigma_{[g]}(z)= \frac{1}{\sqrt{w(N-w)!N!}}\sum_{h \in S_N} \sigma_{hgh^{-1}}(z) \label{eq:gauge invariant sigma symm orb} \ . 
\ee

\begin{figure}
\centering
\begin{tikzpicture}
\draw[thick, red, snake it,bend left=20] (0,0) to (3,2);
\draw[thick, red, snake it,bend left=20] (2,0) to (3,2);
\draw[thick, red, snake it,bend right=20] (6,0) to (3,2);
\fill (3,2) circle (.05) node[above] {$z_*$};
\fill (0,0) circle (.1) node[below] {$\sigma_{g_1}(z_1)$};
\fill (2,0) circle (.1) node[below] {$\sigma_{g_2}(z_2)$};
\node at (4,0) {$\cdots$};
\fill (6,0) circle (.1) node[below] {$\sigma_{g_m}(z_m)$};
\end{tikzpicture}
\caption{The choice or the branch cuts in the correlator $\langle \sigma_{g_1}(z_1) \sigma_{g_2}(z_2) \cdots \sigma_{g_m}(z_m)\rangle$.}\label{fig:correlator ordering}
\end{figure}
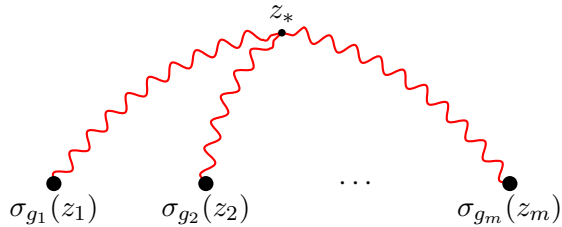

 Let us compute a correlation function of single cycle twist fields of the form \eqref{eq:gauge-inv-corr}. 
Expanding the definition of $\sigma_{[g]}(z)$ \eqref{eq:gauge invariant sigma symm orb}, one can write this as a sum of correlation functions of the (gauge dependent) fields $\sigma_g(z)$, which are themselves independent of $N$. Hence, the $N$ dependence of the correlator is fixed entirely by the combinatorial factors in \eqref{eq:gauge invariant sigma symm orb}. For single cycle twists, we further distinguish gauge dependent correlators \eqref{eq:gauge-dep-corr} by the number of `active colors', meaning the total number of sites involved in all permutations \cite{Pakman:2009zz}. For example, the number of active colors in the correlator
\be 
\langle \sigma_{(1345)}(z_1) \sigma_{(475)}(z_2) \sigma_{(57)}(z_3)\sigma_{(153)}(z_4)\rangle \label{eq:example correlator}
\ee
is 5 (namely the sites 1, 3, 4, 5 and 7 are involved). We will denote the number of active colors by $n$. This is one specific term in the sum of the correlator
\be 
\langle \sigma_4(z_1) \sigma_3(z_2) \sigma_2(z_3) \sigma_3(z_4) \rangle\ .
\ee
Let us fix the ordering of the correlator in which we take successively the OPE, according to \eqref{eq:sigma-sigma-OPE}. We choose the branch cuts as indicated in Figure~\ref{fig:correlator ordering} and hence we take the OPEs in the order in which the fields appear in the correlator. This is the same convention as in \cite{Pakman:2009zz}.\footnote{Notice that this only specifies an ordering up to cyclic permutations. The correlation functions $\langle \sigma_{g_1}(z_1) \cdots \sigma_{g_m}(z_m) \rangle$ are invariant under cyclic permutations. This follows from the fact that cyclically permuted correlators correspond to the same covering map (or the same diagram in the language of \cite{Pakman:2009zz}).} By taking successive OPEs, we see that the correlator is only non-vanishing if
\be 
g_1g_2\cdots g_\nif=\mathds{1}\ .
\label{eq:prod-gs=Id}
\ee
We can restrict the correlation functions further to their connected part. In general, the correlator
\be 
\left \langle \prod_{j=1}^\nif \sigma_{w_j}(z_j) \right \rangle = \left(\prod_{j=1}^\nif \frac{1}{\sqrt{N!(N-w_j)!w_j}}\right) \sum_{\tau_1,\dots \tau_\nif \in S_N} \left \langle \prod_{j=1}^\nif \sigma_{\tau_j(1 \cdots w_j) \tau_j^{-1}}(z_j) \right \rangle\label{eq:gauge dependent correlator SN}
\ee
will also receive contributions from terms, for which the group generated by the group elements
\be 
\left\langle\tau_1(1\cdots w_1)\tau_1^{-1},\, \dots,\, \tau_\nif(1\cdots w_\nif)\tau_\nif^{-1}\right \rangle
\ee
is a direct product of two disjoint subgroups of $S_n$. In this case, the correlator factorizes into the product of correlators of these two subgroups, which hence gives a disconnected contribution. We can conveniently restrict to connected contributions by requiring that the group elements $g_j$ involved in the correlator generate a transitive subgroup of the symmetric group acting on the active colors $S_n$. For instance, this condition is true in the example \eqref{eq:example correlator}. 

As will become apparent momentarily, it is useful to introduce a genus for the individual terms in the sum of eq.~\eqref{eq:gauge dependent correlator SN} by defining
\be 
g\equiv 1-n+\frac{1}{2}\sum_{j=1}^\nif (w_j-1)\ , \label{eq:genus covering map}
\ee
where $w_j$ are the lengths of the involved cycles.
This is always a positive integer, otherwise the correlator \eqref{eq:gauge dependent correlator SN} vanishes.\footnote{Since the identity has even parity, it follows from \eqref{eq:prod-gs=Id} that $\sum_{i=1}^\nif (w_j-1) \in 2\mathds{Z}$. Positivity is implied by the fact that for $g<0$, the subgroup generated by the $g_j$'s cannot be transitive.}
For example, the correlator \eqref{eq:example correlator} is a genus 2 correlator. With these specifications, the correlator \eqref{eq:gauge dependent correlator SN} becomes
\begin{multline} 
\left \langle \prod_{j=1}^\nif \sigma_{w_j}(z_j) \right \rangle_{\!\!\text{c}} =\sum_{g=0}^\infty \left( \prod_{j=1}^\nif \frac{1}{\sqrt{N!(N-w_j)!w_j}} \right) \\
\times\hspace{-1.5cm}\sum_{\genfrac{}{}{0pt}{}{\tau_1,\dots \tau_\nif \in S_N}{\tau_j^{-1} (1 \cdots w_j) \tau_j \text{ generate a transitive subgroup of $S_n$}}} \left \langle \prod_{j=1}^\nif \sigma_{\tau_j^{-1}(1 \cdots w_j) \tau_j}(z_j) \right \rangle\Bigg|_g\ ,\label{eq:gauge dependent correlator SN connected}
\end{multline}
where by the subscript $\text{c}$ we denoted the connected part of the correlator. Actually the sum over $g$ is finite, as follows from \eqref{eq:genus covering map}. The optimal bound on the genus $g$ results from the requirement $n \ge \text{max}_{j=1,\dots,\nif}(w_j)$ \cite{Eberhardt:2019ywk}.

One may ask how many terms in the sum \eqref{eq:gauge dependent correlator SN connected} are actually independent and how many are related by analytic continuation, since by \eqref{eq:sigma-sigma-monodromy}, monodromies of the twist fields will relate different terms in the sum. Let us first account for all trivial combinatorial factors. For each $\tau_j$, there are $|\text{Stab}((1\cdots w_i))|=(N-w_j)!w_j$ trivial choices that do not influence $g_j$. Moreover, there are $\binom{N}{n} $ ways to choose the active colors. After taking care of these factors, we can assume the active colors to be $1$, $\dots,$ $n$ and the genus $g$ correlator becomes
\begin{multline} 
\left \langle \prod_{j=1}^\nif \sigma_{w_j}(z_j) \right \rangle_{\!\!\text{c},g}\!\!\!\! =\binom{N}{n} \left(\prod_{j=1}^\nif \frac{\sqrt{(N-w_j)!w_j}}{\sqrt{N!}} \right) \\
\times\hspace{-1.5cm}\sum_{\genfrac{}{}{0pt}{}{g_j \in [(1 \cdots w_j)] \subset S_n}{g_j\text{ generate transitive subgroup of $S_n$}}} \left \langle \prod_{j=1}^\nif \sigma_{g_j}(z_j) \right \rangle\ ,\label{eq:gauge dependent correlator SN connected sphere}
\end{multline}
where $n=1-g+\tfrac{1}{2}\sum_{j=1}^\nif (w_j-1)$, see eq.~\eqref{eq:genus covering map}. The last factor is now completely $N$-independent and the large $N$ limit behaviour is entirely dictated by the prefactors. There is one last trivial factor which comes from relabelling the active colors $1$, $\dots,$ $n$, which contributes $n!$. Thus, the fundamental correlators one has to compute are classified by  equivalence classes of tuples $(g_1,\dots,g_\nif)$ satisfying:
\begin{enumerate}
\item $g_1,\dots,g_\nif \subset S_n$, and $\langle g_1,\dots,g_\nif \rangle \subset S_n$ is a transitive subgroup.
\item $(g_1,\dots,g_\nif) \sim (g g_1 g^{-1},\dots,g g_\nif g^{-1})$ for any $g \in S_n$. \label{hurwitz item 2}
\item The product of all $g_j$ is $\mathds{1}$.
\end{enumerate}
It is one of the basic facts about \emph{Hurwitz theory}, that this data characterises precisely the possible inequivalent covering maps of the Riemann sphere by a genus $g$ surface with ramification indices $w_j$ at the respective insertion points. The number of such equivalence classes is called the (connected) \emph{Hurwitz number}. For an elementary introduction, see e.g.~\cite{Hurwitzbook}. 

The genus $g$ correlators contribute to \eqref{eq:gauge-inv-corr} at order \cite{Lunin:2000yv, Pakman:2009zz} 
\be 
N^{1-g-\frac{\nif}{2}}=N^{-\frac{1}{2}\chi(g,\nif)}\ ,\label{eq:large N behaviour}
\ee
as can also be seen from applying Stirling's formula to eq.~\eqref{eq:gauge dependent correlator SN connected sphere}.
We rewrote the result in terms of the Euler characteristic of a Riemann surface of genus $g$ with $\nif$ punctures. This suggests that this expansion can be identified with the string worldsheet genus expansion under the AdS/CFT correspondence, with the identification
\be 
g_\text{string}^2\sim \frac{1}{N}\ .
\ee
In particular, the sphere contribution $g=0$ is the leading order for the connected correlator in the large $N$ limit and we shall focus in this paper mostly on sphere correlators.

For the sphere contribution (i.e.~genus 0) to the four-point function, the Hurwitz  number was determined in \cite{Hurwitzmathpaper} and reads
\be 
\min_{j=1,\dots,4}\left(w_j\left(n+1-w_j\right)\right)\ ,\label{eq:4pt Hurwitz number}
\ee
where $n=-1+\tfrac{1}{2}\sum_{j=1}^4 w_j$.
These different coverings or different terms in the sum in eq.~\eqref{eq:gauge dependent correlator SN connected} correspond to the `Feynman diagrams' of \cite{Pakman:2009zz}. 

Until now, we have only taken care of renumbering of terms. We can also make use of analytic continuation. By \eqref{eq:sigma-sigma-monodromy}, tuples 
\be 
(g_1,\dots,g_j,g_{j+1},\dots,g_\nif) \sim (g_1,\dots,g g_j g^{-1},g g_{j+1}g^{-1},\dots,g_\nif)\ , \qquad g=g_jg_{j+1}
\ee
are related by analytic continuation and we hence consider them to be also equivalent (on top of the equivalence relation of \ref{hurwitz item 2}.). We have checked extensively with \texttt{Mathematica} that for the sphere contribution to four-point functions, there is then only one equivalence class for every choice of conjugacy classes $[g_1]$, $\dots,$ $[g_4]$. We have not tried to prove this algebraically, though with sufficient effort this should be possible. By the same token, the different covering maps (counted by \eqref{eq:4pt Hurwitz number}) are all related by analytic continuation in the cross-ratio of the base space. We will see this very explicitly for the example $w_1=w_2=w_3=w_4=2$ in Section~\ref{subsec:2222 covering map}. This fact is known in the mathematical literature as the statement that the Hurwitz space (i.e.~the space of all ramified coverings with given ramification indices of the $n$-punctured sphere) is a connected manifold, see e.g.~\cite[Appendix C]{Hurwitzbook}.

The implication of this statement is that there is only one four-point function to compute at genus 0; all other four-point functions will be related to the basic one via analytic continuation.

\subsection{The covering space method}\label{subsec:covering space method}

The emergence of the genus in correlation functions can be seen much more directly, making manifest the relation of the above equivalence classes of tuples of group elements of $S_n$ to covering maps. The correlation functions \eqref{eq:gauge-dep-corr} of twist operators in symmetric orbifold CFTs were computed by path integral methods in \cite{Lunin:2000yv}. The idea is to pass from the base space (the $z$-space), where according to \eqref{eq:twisted-bondary-cond} the fields have monodromies, to a branched cover $\Sigma_g$ (the $t$-space), where the multiple fields $X^{(k)}(z)$ with monodromies are traded for a single field $X(t)$ without monodromies. The corresponding (branched) covering map
\be 
\Gamma:\ \Sigma_g \longrightarrow \mathbb{CP}^1
\ee
has degree $n$ ($=$ number of active colors). Above, we restricted our interest to connected contributions to the correlation functions, which in the language of the covering space means that $\Sigma_g$ is a connected Riemann surface.

The genus $g$ of the covering surface $\Sigma_g$ is determined by the Riemann-Hurwitz formula, 
\be 
g=1-n+\sum_{j=1}^\nif (w_j-1) \ ,
\ee
where the sum runs over ramification points and $w_i$ are the respective ramification indices. Hence the corresponding genus coincides with \eqref{eq:genus covering map}, thus justifying our notation $\Sigma_g$.

The problem of computing correlation functions of twist fields in the base space can be translated to the problem of computing correlators of (untwisted) fields in the covering space. When transforming the correlator in this way, one has to take into account the conformal factors resulting from the conformal transformation to the covering space. Moreover, the transformation is singular at the insertion points (as well as at spatial infinity, were the curvature of the base space is concentrated). This requires a delicate regularisation process. See \cite{Lunin:2000yv} for more details on this.

\subsection{The covering map}

In the following we will restrict to correlators with genus zero covering space. In this case, the degree of the map is 
\be 
n = \frac{1}{2} \sum_j (w_j -1) + 1 \label{eq:genus zero Riemann Hurwitz}
\ee
and a covering map $\Gamma(t)$ can be computed quite explicitly \cite{Lunin:2000yv, Pakman:2009zz, Roumpedakis:2018tdb}, at least up to four branch points. We focus in the following mostly on the sphere contribution to the four point function
\be 
\left\langle \sigma_{w_1}(0) \sigma_{w_2}(1) \sigma_{w_3}(u) \sigma_{w_4} (\infty)\right \rangle_{\text{c}} .
\ee
As always, we used the conformal Ward identities to fix three of the four coordinates. The remaining cross-ratio is denoted by $u$.

We take the corresponding covering map $\Gamma:\mathbb{CP}^1 \to \mathbb{CP}^1$ to have the expansion
\begin{subequations}\label{eq:Gamma expansions}\
\begin{align}
\Gamma(t)\sim & \ 0 + a_1 t^{w_1}+b_1 t^{w_1+1}+\mathcal{O}(t^{w_1+2}) & \hspace{5pt} t &\sim 0\ , \label{eq:Gamma-at-0}\\
\Gamma(t)\sim & \ 1 + a_2(t-1)^{w_2}+b_2(t-1)^{w_2+1}+\mathcal{O}((t-1)^{w_2+2}) & t &\sim 1\ , \\
\Gamma(t)\sim & \ u + a_3(t-x)^{w_3}+b_3(t-x)^{w_3+1} && \nonumber \\
& \hspace{85pt} + c_3(t-x)^{w_3+2}+ \mathcal{O}((t-x)^{w_3+3})\!\!\! & t &\sim x\ , \\
\Gamma(t)\sim & \ a_4 t^{w_4} + b_4 t^{w_4-1} + \mathcal{O}(t^{w_4-2}) & t &\sim \infty\ .  \label{eq:Gamma-at-inf}
\end{align}
\end{subequations}
We used the automorphism group of $\mathbb{CP}^1$ to also fix three coordinates in the covering space. Thus, there are also four `insertion points' in the covering space at
\be 
t_1=0 \  , \qquad t_2=1 \  , \qquad t_3=x \  , \qquad t_4=\infty \  . 
\ee
Specifying three points in the covering space together with the ramification indices at the branch points determines the covering map up to a finite number of possibilities, which are counted by the respective Hurwitz number. 
As we discussed above, specifying a set of particular representatives in \eqref{eq:gauge dependent correlator SN connected sphere} determines the covering map and the different choices of covering maps correspond to the terms in this sum.
In particular, the covering map depends on the specific cycles involved in the correlation function and not only on their conjugacy classes. 
It is also often useful in the following to change perspective and view $u$, $a_i$ and $b_i$ as (multi-valued) functions of the cross-ratio in the covering space $x$.

The covering map has poles $t =\ell_i$,\footnote{The covering map has of course also a pole at $t=\infty$ by construction, but we do not count $\infty$ as one of the $\ell_i$'s.} where it behaves as 
\be 
\Gamma(t) = \dfrac{C_i}{t- \ell_i} + \mathcal{O}(1) \label{eq:Gamma poles} \  .
\ee
The location of the poles $\ell_i$ as well as their residues $C_i$ are also functions of $x$. The number of poles is easily determined. Poles are the preimages of $\infty$ and there are generically $n$ preimages. $t=\infty$ is however also a preimage of $z=\infty$ and hence there are only $n-w_4$ other poles $\ell_i$.

\section{Differential equation for twisted sector ground state correlators}
\label{sec:diffeq-ssss}

In this section we will show how to find a differential equation for the correlator
\be 
G(u) = \left\langle \sigma_{g_1}(0) \sigma_{g_2}(1) \sigma_{g_3}(u) \sigma_{g_4}(\infty) \right\rangle \  , \label{eq:G correlator}
\ee
where $g_j$ is a single cycle permutation of length $w_j$ for $ j= 1, \dots , 4$ and $\sigma_{g_j}$ has conformal weight $h_j^0$ as in \eqref{eq:conformal-weight-single-cycle}. For notational simplicity, we assume that the active colours in this correlator are the sites $1$, \dots, $n$. We also assume that this correlator gives a connected contribution to the final correlator, since otherwise the computation reduces to a product of two two-point functions.
We work chirally and suppress the dependence on the antiholomorphic coordinate $\bar{u}$. We will reinstate the dependence on $\bar{u}$ in the end.

\begin{figure}
\centering
\begin{tikzpicture}
\draw[thick, red, snake it,bend left=20] (0,0) to (3,2);
\draw[thick, red, snake it,bend left=10] (2,0) to (3,2);
\draw[thick, red, snake it,bend right=10] (4,0) to (3,2);
\draw[thick, red, snake it,bend right=20] (6,0) to (3,2);
\fill (3,2) circle (.05) node[above] {$z_*$};
\fill (0,0) circle (.1) node[below] {$\sigma_{g_1}(0)$};
\fill (2,0) circle (.1) node[below] {$\sigma_{g_3}(u)$};
\fill (4,0) circle (.1) node[below] {$\sigma_{g_2}(1)$};
\fill (6,0) circle (.1) node[below] {$\sigma_{g_4}(\infty)$};
\end{tikzpicture}
\caption{The choice or the branch cuts in the 4-point function.}\label{fig:4 point function branch cuts}
\end{figure}
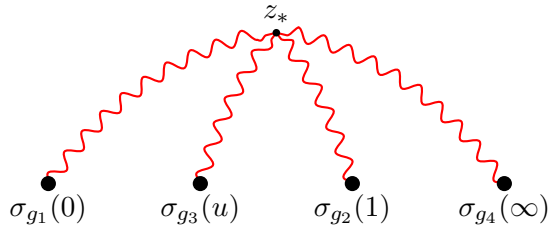

In this section, we will assume a slightly different ordering of the twist fields. We choose the branch cuts as in Figure~\ref{fig:4 point function branch cuts}. Hence the correlator is only non-vanishing if $g_1g_3g_2g_4=1$. We do this to facilitate taking the limit $u\to 0$ in the following Section.
\subsection{Turning null vectors into differential equations}
\label{sec:null-vectors-into-differential-equations}
Consider the quantity 
\be 
\oint_C \mathrm{d}z\  \left\langle \sum_{k=1}^{n} f_k(z) T^{(k)}(z) \sigma_{g_1}(0) \sigma_{g_2}(1) \sigma_{g_3}(u) \sigma_{g_4}(\infty) \right\rangle \  ,  
\label{eq:Fsigmas}
\ee 
where $C$ is a closed path around the insertion points and $f_k(z)$ are multivalued functions obeying the following properties (we will address their existence in Section~\ref{subsec:choosing-f})
\begin{enumerate}
\item Close to the insertion points $z_j$, for $k = 1, \dots, n$  \label{item 1}
\be 
f_{k}(e^{2 \pi i}z +z_j) = f_{g_j(k)}(z+z_j) \  , \qquad j = 1,\dots,4 \  .  
\label{eq:f-monodromy}
\ee 
\item $f_k(z)$ has no poles except at infinity, where the behaviour is specified through item~\ref{item 3}.\label{item 2}
\item Close to the insertion points $z_j$, for $k = 1, \dots, n$, the functions $f_k$ have expansions
\begin{subequations}
\begin{align}
f_k(z) &=  \gamma_1 z + \mathcal{O}\left(z^{\frac{w_1+1}{w_1}}\right) \ , \qquad && z \sim 0 \ , \label{eq:f-expansion 0} \\
f_k(z) &=  \gamma_2(z-1) + \mathcal{O}\left((z-1)^{\frac{w_1+1}{w_1}}\right)\ , \qquad && z \sim 1 \ , \label{eq:f-expansion 1} \\
f_k(z) &= \alpha_3 + \beta_3 \mathrm{e}^{-\frac{2\pi i k}{w_3}} (z-u)^{\frac{w_3-1}{w_3}} + \gamma_3(z-u) && \nonumber \\
& \hspace{60pt} +  \delta_3 \mathrm{e}^{\frac{2\pi i k}{w_3}} (z-u)^{\frac{w_3+1}{w_3}} + \mathcal{O}\left((z-u)^{\frac{w_3+2}{w_3}}\right) \ , \qquad && z \sim u \ ,  \label{eq:f-expansion u} \\
f_k(z) &=  \gamma_4 z+\mathcal{O}\left(z^{\frac{w_4-1}{w_4}}\right)  \qquad && z \sim \infty \ ,\label{eq:f-expansion-inf}
\end{align}
\end{subequations}
where in \eqref{eq:f-expansion u} we made explicit the $k$-dependence due to the $w_3$ different branches.  The term $\delta_3$ will only be needed in Section~\ref{sec:more-differential-equations}.\label{item 3}
\end{enumerate}
Item \ref{item 1} specifies the monodromy properties of the functions $f_k(z)$. They are designed to cancel the monodromies of the correlation function inherited from \eqref{eq:monodromy T}. Hence, \ref{item 1} makes the entire integrand in \eqref{eq:Fsigmas} single-valued in $z$. Item \ref{item 2} ensures that the integrand does not have any poles away from the insertion points.
Thus, the integrand in \eqref{eq:Fsigmas} is single-valued and has no poles outside of the integration contour $C$ and therefore vanishes, 
\be 
\oint_C \mathrm{d}z\ \left\langle  \sum_{k=1}^{n} f_k(z) T^{(k)}(z) \sigma_{g_1}(0) \sigma_{g_2}(1) \sigma_{g_3}(u) \sigma_{g_4}(\infty) \right\rangle = 0 \  . 
\label{eq:Fsigmas=0}
\ee 
Inside the contour $C$, there are poles at the insertion points. We will now evaluate the contribution arising from poles at $0,1,u$ and $\infty$ in turn. Let us start from $u$. Since $T^{(k)}$ has regular OPE with $\sigma_{g_3}$ for $k \notin  g_3$ (by $k \in g$ we mean that $k$ is an active colour in $g$), the contribution from the pole at $z=u$ can be rewritten as 
\be 
\oint_u \mathrm{d}z\ \left\langle \sigma_{g_1}(0) \sigma_{g_2}(1)  \sum_{k\in g_3} f_k(z) T^{(k)}(z) \sigma_{g_3}(u) \sigma_{g_4}(\infty) \right\rangle \  . 
\ee
Without loss of generality, we can assume that $g_3 = (1,2,\dots, w_3)$ and rewrite
\begin{align} 
\sum_{k=1}^{w_3} f_k(z) T^{(k)}(z) &= \frac{1}{w_3} \sum_{\ell=0}^{w_3 -1} \sum_{k=1}^{w_3} f_k(z) \mathrm{e}^{\frac{2 \pi i k \ell}{w_3}} \sum_{r=1}^{w_3} T^{(r)} \mathrm{e}^{-\frac{2 \pi i r \ell}{w_3} }\\
&= \frac{1}{w_3} \sum_{\ell=0}^{w_3 -1} \sum_{k=1}^{w_3} f_k(z) \mathrm{e}^{\frac{2 \pi i k \ell}{w_3}}  T^{\ell} (z) \ ,
\label{eq:F-sum}
\end{align}
where we made use of the Fourier transformed generators \eqref{eq:Fourier transformed Ts}.
The contribution from the pole at $u$ is then 
\begin{multline} \label{eq:diff-eq-derivation1}
\oint_u \mathrm{d}z\ \left\langle \sigma_{g_1}(0) \sigma_{g_2}(1)  \sum_{k\in g_3} f_k(z) T^{(k)}(z) \sigma_{g_3}(u) \sigma_{g_4}(\infty) \right\rangle  \\
\hspace{-35pt} = \oint_{u} \mathrm{d}z\  \sum_{\ell=0}^{w_3-1} \sum_{k=1}^{w_3} \dfrac{\mathrm{e}^{\frac{2 \pi i k \ell}{w_3}} f_k(z)}{w_3}  \Bigg(\dfrac{\left\langle \sigma_{g_1}(0) \sigma_{g_2}(1)  (L_{-\ell/w_3}\sigma_{g_3})(u) \sigma_{g_4}(\infty) \right\rangle}{(z-u)^{2-\frac{\ell}{w_3}}} \\
 + \dfrac{\left\langle \sigma_{g_1}(0) \sigma_{g_2}(1) (L_{-1-\ell/w_3} \sigma_{g_3})(u) \sigma_{g_4}(\infty) \right\rangle}{(z-u)^{1-\frac{\ell}{w_3}}}  \Bigg) \  , 
\end{multline}
where we have used the OPE \eqref{eq:OPE-Tsigma} in the second equality. We can now insert the expansion of $f_k(z)$ around $u$, as specified by item \ref{item 3}. We should note that once we insert the expansion of $f_k(z)$, there are only three terms in total contributing. The contribution from $z=u$ becomes
\begin{multline}
\oint_u \mathrm{d}z\ \left\langle \sigma_{g_1}(0) \sigma_{g_2}(1)  \sum_{k\in g_3} f_k(z) T^{(k)}(z) \sigma_{g_3}(u) \sigma_{g_4}(\infty) \right\rangle \\
=\left(\alpha_3 \partial_{u} + h_3^0 \gamma_3\right)\left\langle \sigma_{g_1}(0) \dots \sigma_{g_4}(\infty) \right\rangle  + \beta_3 \left\langle  \sigma_{g_1}(0) \sigma_{g_2}(1) (L_{-\frac{1}{w_3}}\sigma_{g_3})(u) \sigma_{g_4}(\infty) \right\rangle\ .
\label{eq:diff-eq-derivation2}
\end{multline}
Finally, we use the property that $(L_{-1/w_3}\sigma_{g_3})(u)$ is a null field and hence the last correlation function vanishes.
We simply end up with the contribution 
\begin{multline}
\oint_u \mathrm{d}z\ \left\langle \sigma_{g_1}(0) \sigma_{g_2}(1)  \sum_{k\in g_3} f_k(z) T^{(k)}(z) \sigma_{g_3}(u) \sigma_{g_4}(\infty) \right\rangle \\
=\left(\alpha_3 \partial_{u} + h_3^0 \gamma_3\right)\left\langle \sigma_{g_1}(0)\sigma_{g_2}(1)\sigma_{g_3}(u)  \sigma_{g_4}(\infty) \right\rangle \ 
\label{eq:diff-eq-derivation3}
\end{multline}
from $z=u$.

The contribution from the other poles can be computed analogously. The only minor difference arises when computing the integral around $z=\infty$, since the integral runs clockwise viewed from $\infty$, instead of counterclockwise. This leads to an additional minus sign. Hence we obtain the differential equation
\begin{multline} 
\oint_C \mathrm{d}z\  \left\langle \sum_{k=1}^{n} f_k(z) T^{(k)}(z) \sigma_{g_1}(0) \sigma_{g_2}(1) \sigma_{g_3}(u) \sigma_{g_4}(\infty) \right\rangle \\
=\left(\alpha_3 \partial_u+(h_1^0 \gamma_1 + h_2^0 \gamma_2 + h_3^0 \gamma_3 - h_4^0 \gamma_4)\right)\left\langle \sigma_{g_1}(0)\sigma_{g_2}(1)\sigma_{g_3}(u)  \sigma_{g_4}(\infty) \right\rangle=0\ .\label{eq:diff-eq-ssss-1}
\end{multline}
We also note that we have actually only made use of the null vector $L_{-1/w_3}\sigma_{g_3}=0$. Therefore this differential equation is actually valid for arbitrary primary fields $\mathcal{O}_{g_1}(0)$, $\mathcal{O}_{g_2}(1)$ and $\mathcal{O}_{g_4}(\infty)$ in the respective twisted sector. Only $\sigma_{g_3}(u)$ is required to be the bare twist field. We explore this possibility further in Section~\ref{subsec:one twist field}. For the moment we consider the case where all four fields are bare twist fields.

\subsection[Constructing the functions \texorpdfstring{$f_k$}{fk}]{Constructing the functions $\boldsymbol{f_k}$}
\label{subsec:choosing-f}
While we have derived a differential equation for the correlator, we still have to demonstrate that there exist functions $f_k$ satisfying items~\ref{item 1}, \ref{item 2} and \ref{item 3}. Moreover, we have to express the coefficients (which are themselves functions of $u$) in terms of known data.

We can satisfy item \ref{item 1} by writing
\be 
f_k=h \circ \Gamma_k^{-1}\ ,
\label{eq:f_k=h-comp-Gamma-minus1}
\ee
where we have denoted by $\{\Gamma^{-1}_k\}_{k=1,\dots, n}$ the $n$ inverses of the covering map $\Gamma(t)$ that corresponds to the particular representatives of the conjugacy classes of the correlator \eqref{eq:G correlator}. Moreover, $h$ is a (single-valued) meromorphic function on the covering space.

It turns out that setting 
\be 
h=\partial_x\Gamma
\label{eq:h=dixGamma}
\ee
satisfies also items \ref{item 2} and \ref{item 3}.\footnote{Remember that the covering map is a function of $t$, but it also depends on the cross-ratio $x$ in the covering space.}
For it, notice that $\partial_x \Gamma$ can only have poles where $\Gamma$ has poles, i.e.~at $t = \ell_i$ or $t=\infty$. Hence $ \partial_x \Gamma \left(\Gamma^{-1}_k(z) \right)$ has a pole at $z$ if $\Gamma_k^{-1}(z)=\ell_i$ or $\infty$, which implies $z=\infty$. Hence $ \partial_x \Gamma \left(\Gamma^{-1}_k(z) \right)$ has only a pole at $z=\infty$.
 
The corresponding values for $\alpha_3$ and $\gamma_j$ can be expressed in terms of the covering map data \eqref{eq:Gamma expansions} as 
\begin{subequations} \label{eq:fk parameters}
\begin{align}
\alpha_3 &= \partial_x u \  ,  \label{eq:alphas}\\
 \gamma_j  &= \frac{\partial_x a_j}{a_j} \  , \  j \neq 3 \  , \\
  \gamma_3 &= \frac{\partial_x a_3 - 2 b_3}{a_3} \label{eq:gammas}\  .   
\end{align}
\end{subequations}

\subsection{Solving the differential equation}

Inserting equations \eqref{eq:fk parameters} into the differential equation \eqref{eq:diff-eq-ssss-1} we find 
\begin{multline}
\Biggl[ \partial_x + \dfrac{c}{24} \sum_{j=1}^3 \left(\frac{w_r^2-1}{w_r} \right) \frac{\partial_x a_r}{a_r} - \frac{c}{24}\left(\frac{w_4^2-1}{w_4} \right) \frac{\partial_x a_4}{a_4} \\
- \frac{c}{12} \left(\frac{w_3^2-1}{w_3} \right) \frac{b_3}{a_3} \Biggr] G(x) = 0 \  , 
\label{eq:diff-eq-ssss-2}
\end{multline}
where we have written the correlator \eqref{eq:G correlator}
as a function of $x$. Indeed, $u$ depends on $x$ through $\Gamma(x)=u$.  
In Appendix~\ref{app:Gamma-identities} we prove the identity 
\be 
\sum_{j=1}^3 (w_j-1)\frac{\partial_x a_j}{a_j}-(w_4-1)\frac{\partial_x a_4}{a_4}- \frac{2b_3(w_3^2-1)}{a_3w_3}-2\sum_{i} \frac{\partial_x C_i}{C_i}=0 
\ee
between the parameters  \eqref{eq:Gamma expansions} and \eqref{eq:Gamma poles} of the covering map.
The differential equation~\eqref{eq:diff-eq-ssss-2} can thus be rewritten as 
\begin{equation}
\left[ \partial_x + \dfrac{c}{24} \sum_{j=1}^3 \left(\frac{w_j-1}{w_j} \right) \frac{\partial_x a_j}{a_j} - \frac{c}{24}\left(\frac{w_4-1}{w_4} \right) \frac{\partial_x a_4}{a_4} + \frac{c}{12} \sum_{i}\frac{\partial_x C_i}{C_i} \right] G(x) = 0 \  . 
\label{eq:diff-eq-ssss-3}
\end{equation}
It is now trivial to find the solution to this equation, which is given by
\be 
G(x) = C_{w_1,w_2,w_3,w_3} \ a_1^{-\frac{c(w_1-1)}{24w_1}} a_2^{-\frac{c(w_2-1)}{24w_2}} a_3^{-\frac{c(w_3-1)}{24w_3}} a_4^{\frac{c(w_4-1)}{24w_4}} \prod_i C_i^{-\frac{c}{12}} \  , 
\ee
where $C_{w_1,w_2,w_3,w_4}$ is an arbitrary integration constant. We have of course an analogous antiholomorphic differential equation and by putting together holomorphic and antiholomorphic contributions, we finally find
\begin{multline}
G_{g_1,g_2,g_3,g_4}(x,\bar{x}) = \left| C_{w_1,w_2,w_3,w_4}\right|^2 \ |a_1|^{-\frac{c(w_1-1)}{12w_1}} |a_2|^{-\frac{c(w_2-1)}{12w_2}} \times \\
\times  |a_3|^{-\frac{c(w_3-1)}{12w_3}} |a_4|^{\frac{c(w_4-1)}{12w_4}} \prod_i |C_i|^{-\frac{c}{6}} \  , \label{eq:solution diff eq}
\end{multline}
in agreement with the results of \cite{Lunin:2000yv} and \cite{Pakman:2009zz}. We remind the reader that the right-hand side is still gauge dependent on the choices of representatives $g_1$, \dots, $g_4$, since it depends on the specific choice of the covering map. However, as we have discussed in Section~\ref{sec:symm orbifold}, all sphere four-point functions with different choices of $g_j$ are related by analytic continuation in $x$. Because of this, the integration constant $C_{w_1,w_2,w_3,w_3}$ can only depend on the conjugacy class and not on the representatives. This will be confirmed below by the explicit calculation of $C_{w_1,w_2,w_3,w_4}$.

The differential equation \eqref{eq:diff-eq-ssss-3} alone is not able to predict the integration constant $C_{w_1,w_2,w_3,w_4}$. However, we will show in the next section that it can be fixed by imposing the factorisation of the four-point function into the product of two three-point functions.

\section{Fixing the integration constant and factorisation}\label{sec:fixing integration constant}
In this section, we analyse the factorisation of the (gauge dependent) four-point function
\be 
G_{g_1,g_2,g_3,g_4}(u,\bar{u})=\left\langle \sigma_{g_1}(0) \sigma_{g_2}(1) \sigma_{g_3}(u) \sigma_{g_4}(\infty) \right\rangle \ . \label{eq:gauge-dependent-4pt}
\ee
By requiring that the four-point function factorises into the product of two three-point functions, we will be able to determine the integration constant in \eqref{eq:solution diff eq}. 

We impose factorisation directly for \eqref{eq:gauge-dependent-4pt} into three-point functions as follows
\be 
G_{g_1, g_2, g_3, g_4}(u,\bar{u})=G_{g_1, g_3, 1, (g_1g_3)^{-1}} \ G_{g_1g_3,g_2, 1, g_4} \ |u|^{-2 h_{w_1}^0-2 h_{w_3}^0 + 2 h_{w}^0  }+\cdots\ . \label{eq:factorisation-1-channel}
\ee 
Here, we have denoted three-point functions by the same symbol as the corresponding four-point functions, where the third field is the identity field. Because of this, they do not depend on $x$ and we omit the argument. In this factorisation, $w$ is the length of the single cycle $g_1g_3$. By assumption, the covering space is a sphere, which implies that only single cycle fields run in the OPE channel. This is the requirement that the 4-point function obeys the OPE \eqref{eq:sigma-sigma-OPE}. Note that we are taking the OPE consistent with our ordering convention of Figure~\ref{fig:4 point function branch cuts}.

In the following, for notation simplicity, we will fix $g_1, \dots, g_4$ (or equivalently up to relabelling of the active colors one covering space together with a covering map) and simply write $G(x,\bar{x})$ for $G_{g_1,g_2,g_3,g_4}(x,\bar{x})$. In Section~\ref{subsec:determining normalisation factors} below, by imposing eq.~\eqref{eq:factorisation-1-channel}, we will deduce a recursion relation on the so far unrestricted integration constants. This will in turn permit to fix them uniquely. 

\subsection{An explicit parametrisation of the covering map}\label{subsec:parametrisation covering map}

In order to proceed, it will be important to understand how the coefficients $a_j$ and $C_i$ behave in the limit $u \to 0$. For this purpose, it is useful to employ an explicit parametrisation of the covering map. We choose the parametrisation advocated in \cite{Roumpedakis:2018tdb}. The idea is to determine the derivative of the covering map $\partial \Gamma(t)$. The derivative has a zero of order $w_j-1$ at $t_j$ (and nowhere else) and hence we can write
\be 
\partial \Gamma(t)=\mathcal{N}\ \frac{t^{w_1-1}(t-1)^{w_2-1}(t-x)^{w_3-1}}{\prod_{i=1}^{n-w_4} (t-\ell_i)^2}\ .
\ee 
Remember that this is a sphere correlator and hence $n$ is given by \eqref{eq:genus zero Riemann Hurwitz}.
The locations of the poles are determined by requiring that $\partial \Gamma(t)$ is a total derivative and hence all its residues vanish, $\mathop{\text{Res}}_{t=\ell_i} \partial \Gamma(t)=0$. This implies
\be 
\frac{w_1-1}{\ell_i}+\frac{w_2-1}{\ell_i-1}+\frac{w_3-1}{\ell_i-x}-2\sum_{j \ne i} \frac{1}{\ell_i-\ell_j}=0\ .\label{eq:equations l}
\ee
These $n-w_4$ equations determine a finite number of solution for the $n-w_4$ unknowns. The constant $\mathcal{N}$ is fixed by requiring that
\be 
1=\Gamma(1)=\int_0^1 \mathrm{d}t\ \partial \Gamma(t)\ ,
\ee
and hence
\be 
\mathcal{N}=\left(\int_0^1 \mathrm{d}t\ \frac{t^{w_1-1}(t-1)^{w_2-1}(t-x)^{w_3-1}}{\prod_{i=1}^{n-w_4} (t-\ell_i)^2}\right)^{-1}\ .
\ee
As noted in \cite{Roumpedakis:2018tdb}, it is convenient to rewrite the solution \eqref{eq:solution diff eq} in terms of these variables, where it takes the form
\begin{align}
G(u,\bar{u})&=\frac{D_{w_1,w_2,w_3,w_4}}{|\mathcal{N}|^{2(\Delta_1+\Delta_2+\Delta_3-\Delta_4)}}|x|^{-\frac{c(w_1+w_3)(w_1-1)(w_3-1)}{12w_1w_3}} |1-x|^{-\frac{c(w_2+w_3)(w_2-1)(w_3-1)}{12w_2w_3}}\nonumber\\
&\qquad\times\prod_i |\ell_i|^{-\frac{c(w_1-1)^2}{6w_1}}|\ell_i-1|^{-\frac{c(w_2-1)^2}{6w_2}}|\ell_i-x|^{-\frac{c(w_3-1)^2}{6w_3}}\prod_{i\ne j} |\ell_i-\ell_j|^{\frac{c}{3}}\ ,\label{eq:four point function l parametrisation}
\end{align}
with
\be
D_{w_1,w_2,w_3,w_4}=|C_{w_1,w_2,w_3,w_4}|^2 w_1^{\frac{c(w_1-1)}{12w_1}}w_2^{\frac{c(w_2-1)}{12w_2}}w_3^{\frac{c(w_3-1)}{12w_3}}w_4^{-\frac{c(w_4-1)}{12w_4}}\ . \label{eq:CD relation}
\ee
\subsection[The limit \texorpdfstring{$u\to 0$}{u->0}]{The limit $\boldsymbol{u\to 0}$}\label{subsec:limit x0}

To discuss factorisation, it is important to understand the behaviour of the covering map as $u$ approaches the other insertion points. Let us analyse the limit $u \to 0$ and hence $x \to 0$ of the system of equations \eqref{eq:equations l} determining the $\ell_i$'s.

There are two possible behaviours of $\ell_i$ in the limit $x\to 0$. We can either have
\be 
\ell_i=\mathcal{O}(x) \qquad \text{or}\qquad  \ell_i=\mathcal{O}(1) \ .
\ee
No other scalings are possible.
By permutation symmetry of the equations, we will hence assume without loss of generality that
\begin{align}
\ell_i&=x \ell^{(1)}_i+\mathcal{O}(x^2)\ ,  & & i=1,\dots,k_1\ , \\
\ell_i&=\ell_i^{(2)}+\mathcal{O}(x)\ , & & i=k_1+1,\dots,k_1+k_2=n-w_4\ .
\end{align}
The two groups $\ell_i^{(1)}$ and $\ell_i^{(2)}$ satisfy in the limit separately equations of the form
\begin{align}
0&=\frac{w_1-1}{\ell_i^{(1)}}+\frac{w_3-1}{\ell_i^{(1)}-1}-2\sum_{j \ne i} \frac{1}{\ell_i^{(1)}-\ell_j^{(1)}}\ , \\
0&=\frac{w_1+w_3-2k_1-2}{\ell_i^{(2)}}+\frac{w_2-1}{\ell_i^{(2)}-1}-2\sum_{j \ne i} \frac{1}{\ell_i^{(2)}-\ell_j^{(2)}}\ ,
\end{align}
which are the equations determining the locations of the poles for the four-point functions with parameters
\begin{subequations}
\begin{align}
w_1^{(1)}&=w_1 \  , & w_2^{(1)}&=w_3 \  , & w_3^{(1)}&=1 \  , & w_4^{(1)}&=w_1+w_3-2k_1-1 \  , \\
w_1^{(2)}&=w_1+w_3-2k_1-1 \  , & w_2^{(2)}&=w_2 \  , & w_3^{(2)}&=1 \  , & w_4^{(2)}&=w_4 
\end{align}\label{eq:w_i}
\end{subequations}
respectively. Of course, since $w_3^{(1)}=w_3^{(2)}=1$, these four-point functions have one identity operator inserted and are hence actually three-point functions. For consistency, we stick with our conventions and continue to label them as four-point functions.

With these preparations, it is also straightforward to extract the leading behaviour of other quantities. We have 
\begin{align}
\mathcal{N}^{-1}&=\int_0^1 \mathrm{d}t\ \frac{t^{w_1-1}(t-1)^{w_2-1}(t-x)^{w_3-1}}{\prod_{i=1}^{n-w_4} (t-\ell_i)^2}\\
&=\int_0^1  \mathrm{d}t\ \frac{t^{w_1+w_3-2k_1-2}(t-1)^{w_2-1}}{\prod_{i=k_1+1}^{n-w_4} (t-\ell_i)^2}+\mathcal{O}(x)\\
&=(\mathcal{N}^{(2)})^{-1}+\mathcal{O}(x)\ ,
\end{align}
where we have used the hopefully intuitive notation that $\mathcal{N}^{(2)}$ is the normalisation constant of the four-point function defined by $w_j^{(2)}$, as specified in \eqref{eq:w_i}. We also need the behaviour of the cross-ratio $u=\Gamma(x)$ as $x\to 0$, for which we compute
\begin{align}
u&=\mathcal{N} \int_0^x \mathrm{d}t\ \frac{t^{w_1-1}(t-1)^{w_2-1}(t-x)^{w_3-1}}{\prod_{i=1}^{n-w_4} (t-\ell_i)^2}\\
%&=\mathcal{N} x^{w_1+w_3-1} \int_0^1 \mathrm{d}t\ \frac{t^{w_1-1}(t-1)^{w_3-1}(tx-1)^{w_2-1}}{\prod_{i=1}^{k_1} (xt-x\ell_i^{(1)})^2\prod_{i=k_1+1}^{n-w_4} (xt-\ell_i^{(2)})^2}\\
&=\mathcal{N}^{(2)}\int_0^1 \mathrm{d}t\ \frac{ (-1)^{w_2+1}x^{w_1+w_3-1-2k_1} t^{w_1-1}(t-1)^{w_3-1}}{\prod_{i=1}^{k_1} (t-\ell_i^{(1)})^2\prod_{i=k_1+1}^{n-w_4} (-\ell_i^{(2)})^2}+\mathcal{O}(x^{w_1+w_3-2k_1})\\
&=\frac{\mathcal{N}^{(2)}}{\mathcal{N}^{(1)}} (-1)^{w_2+1}x^{w_1+w_3-1-2k_1} \prod_{i=k_1+1}^{n-w_4} (-\ell_i^{(2)})^{-2}+\mathcal{O}(x^{w_1+w_3-2k_1})\ .
\end{align}

\subsection{Factorisation of the four-point function}\label{subsec:factorisation}
After these preliminaries, we can determine the behaviour of $G(x,\bar{x})$ in the limit $x\to 0$. For this, we use the parametrisation \eqref{eq:four point function l parametrisation}. Let us denote by $G_{w_1, w_3, 1, w}$ and by $G_{w,w_2, 1, w_4}$ the four-point functions with twists $w_j^{(1)}$ and $w_j^{(2)}$ -- they do not depend on $x$, since ${w_3^{(1)}=w_3^{(2)}=1}$. This again reflects the fact that they effectively are three-point functions. We have $w=w_4^{(1)}=w_1^{(2)}=w_1+w_3-2k_1-1$, see eqs.~\eqref{eq:w_i}. Then a direct computation yields
\begin{align}
G(u,\bar{u})=\frac{D_{w_1,w_2,w_3,w_4}}{D_{w_1,w_3,1,w}D_{w,w_2,1,w_4}} G_{g_1, g_3, 1, (g_1g_3)^{-1}} G_{g_1g_3,g_2, 1, g_4} |u|^{-2h_{w_1}^0-2h_{w_3}^0+2h_{w}^0}+\cdots\ ,
\end{align}
where the dots stand for higher order terms in $u$. $g_1g_3$ is the single-cycle of length $w$ in which $\sigma_{g_1}(0)$ and $\sigma_{g_3}(u)$ fuse, see eq.~\eqref{eq:sigma-sigma-OPE}.

This is exactly the correct contribution of the OPE channel
\be 
\begin{tikzpicture}[baseline={([yshift=-.5ex]current bounding box.center)}]
\draw[thick] (-1,1.73) to node[left] {$w_1$} (0,0)  to node[above] {$w$} (2,0) to node[right] {$w_2$} (3,1.73);
\draw[thick] (-1,-1.73) to node[left] {$w_3$} (0,0);
\draw[thick] (3,-1.73) to node[right] {$w_4$} (2,0);
\end{tikzpicture}
\ee
Notice that different fusion channels correspond to different values of ${k_1 = 0, \dots, (n-w_4)}$. We should note that $k_1$ was determined by choosing a particular branch of solutions of \eqref{eq:equations l} and hence a particular covering space. Since the covering space in turn is determined by the choice of group elements $g_1$, \dots, $g_4$, $k_1$ is implicitly specified by the group elements. By comparing with \eqref{eq:factorisation-1-channel}, we find
\be 
D_{w_1,w_2,w_3,w_4}=D_{w_1,w_3,1,w}D_{w,w_2,1,w_4}\label{eq:factorisation four point function}
\ee
for any $w$ for which the two sides are non-vanishing.
\subsection{Determining the normalisation factors} \label{subsec:determining normalisation factors}
Now that we have derived \eqref{eq:factorisation four point function}, it is easy to deduce their general form. We first compute explicitly $D_{w,w,1,1}$ by requiring that the two-point function is unit normalised. In this case, the covering map is given by
\be 
\Gamma(t)=\frac{t^w}{t^w-(t-1)^w}\ ,
\ee
and hence everything can be computed explicitly. We obtain
\be 
D_{w,w,1,1}=w^{-\frac{c(w^2+1)}{6}}\ .
\ee
An even easier calculation gives also
\be 
D_{w,1,1,w}=1\ ,
\ee
since in this case the covering map is simply given by $\Gamma(t)=t^w$.\footnote{Note that the definition of $D_{w_1,w_2,w_3,w_4}$ is not completely symmetric, since we treated the field at infinity differently. As advocated in \cite{Eberhardt:2019ywk}, one can give a more symmetric definition by defining $a_4^\text{their}=(-1)^{w_4+1} (a_4^\text{our})^{-1}$, which would remove various minus signs. We stick with the conventions of \cite{Pakman:2009zz, Roumpedakis:2018tdb}, which results in slightly asymmetric formulas. \label{footnote:a4}}

Since $G(x,\bar{x})$ is a bosonic four-point function, it enjoys a permutation symmetry in the insertion points $0$, $1$ and $x$, provided that we exchange simultaneously also $w_1$, $w_2$ and $w_3$.\footnote{It is straightforward to prove this also directly from our parametrisation \eqref{eq:four point function l parametrisation}, by applying simultaneously a M\"obius transformation of the covering map both in the covering space and in the base space.} Since our parametrisation discriminates the point at $\infty$, we will not use permutation symmetry with the fourth insertion point. $D_{w_1,w_2,w_3,w_4}$ is hence symmetric in the first three arguments. Following these steps, one can compute the most general $D_{w_1,w_2,w_3,w_4}$ recursively:
\begin{enumerate}
\item \label{item:derivation D 1}Compute $D_{w_1,w_2,w_1,w_2}$:
\be 
D_{w_1,w_2,w_1,w_2}=D_{w_1,w_1,1,1}D_{1,w_2,1,w_2}=w_1^{-\frac{c(w_1^2+1)}{6}}\ .
\ee
\item Next, one derives the relation between $D_{w_1,w_2,1,w_3}$ and $D_{w_3,w_2,1,w_1}$:
\begin{align}
D_{w_1,w_2,1,w_3}=D_{w_1,w_1,w_2,w_2} D_{w_1,w_3,1,w_2}^{-1}
%&=D_{w_1,w_1,w_2,w_2} D_{w_3,w_3,w_1,w_1}^{-1} D_{w_2,w_3,1,w_1}\\
=w_1^{-\frac{c(w_1^2+1)}{6}}w_3^{\frac{c(w_3^2+1)}{6}}D_{w_2,w_3,1,w_1}\ .
\end{align}
\item This leads to the three-point functions:
\begin{align}
w_1^{-\frac{c(w_1^2+1)}{6}}
%= D_{w_1,w_1,w_3,w_3} 
=D_{w_1, w_3,1,w_2} D_{w_2,w_1,1,w_3} 
= w_2^{\frac{c(w_2^2+1)}{6}}w_3^{-\frac{c(w_3^2+1)}{6}} (D_{w_1,w_2,1,w_3})^2\ ,
\end{align}
and hence\footnote{We choose a positive sign. This sign is not yet fixed by the normalisation and can be chosen freely.}
\be 
D_{w_1,w_2,1,w_3}=w_1^{-\frac{c(w_1^2+1)}{12}}w_2^{-\frac{c(w_2^2+1)}{12}}w_3^{\frac{c(w_3^2+1)}{12}}\ .
\ee
\item It now follows immediately from the factorisation \eqref{eq:factorisation four point function} that
\be 
D_{w_1,w_2,w_3,w_4}=w_1^{-\frac{c(w_1^2+1)}{12}}w_2^{-\frac{c(w_2^2+1)}{12}}w_3^{-\frac{c(w_3^2+1)}{12}}w_4^{\frac{c(w_4^2+1)}{12}}\ .\label{eq:solution D}
\ee
This solution satisfies all the contraints we imposed and is hence the unique solution.
\end{enumerate}
Expressed in terms of the original constants $C_{w_1,w_2,w_3,w_4}$, we have \eqref{eq:CD relation}
\be 
|C_{w_1,w_2,w_3,w_4}|^2=w_1^{-\frac{c(w_1+1)}{12}}w_2^{-\frac{c(w_2+1)}{12}} w_3^{-\frac{c(w_3+1)}{12}} w_4^{\frac{c(w_4+1)}{12}}\ .
\ee
Hence the final result for the four-point function of twisted sector ground states is
\begin{multline}
\left\langle \sigma_{g_1}(0) \sigma_{g_2}(1)  \sigma_{g_3}(u) \sigma_{g_4}(\infty) \right\rangle  = w_1^{-\frac{c(w_1+1)}{12}}w_2^{-\frac{c(w_2+1)}{12}} w_3^{-\frac{c(w_3+1)}{12}} w_4^{\frac{c(w_4+1)}{12}} |a_1|^{-\frac{c(w_1-1)}{12w_1}} \\
\times|a_2|^{-\frac{c(w_2-1)}{12w_2}} |a_3|^{-\frac{c(w_3-1)}{12w_3}} |a_4|^{\frac{c(w_4-1)}{12w_4}} \prod_j |C_j|^{-\frac{c}{6}}\ .\label{eq:final result four point function}
\end{multline}
This coincides with the result of \cite{Lunin:2000yv} and \cite{Pakman:2009zz}.

\section{An example: The four-point function \texorpdfstring{$\boldsymbol{\langle \sigma_2(0)\sigma_2(1)\sigma_2(u)\sigma_2(\infty)\rangle}$}{<sigma2 sigma2 sigma2 sigma2>}}\label{sec:example}

To exemplify the above computation at a simple instance, we compute the four-point function
\be 
\langle \sigma_2(0)\sigma_2(1)\sigma_2(u)\sigma_2(\infty)\rangle \label{eq:gauge-dep-corr-2222}
\ee
explicitly. In this case, the covering map can be either of degree 3 or 2, and the covering space has genus 0 or 1, respectively (see \eqref{eq:genus covering map}). We will focus on the genus $0$ contribution. Before computing this, let us first comment on the disconnected part of the correlator. 

\subsection{The disconnected contribution}
There is also a disconnected contribution which we shall discuss first for completeness. The disconnected covering space comes from the terms (up to relabelling of the active colors)
\begin{subequations}
\begin{align}
&\langle \sigma_{(12)}(0) \sigma_{(12)}(1) \sigma_{(34)}(u) \sigma_{(34)}(\infty) \rangle\nonumber\\
&\qquad\qquad\qquad=\langle \sigma_{(12)}(0) \sigma_{(12)}(1)\rangle\langle \sigma_{(34)}(u) \sigma_{(34)}(\infty)\rangle= 1 \ , \\
&\langle \sigma_{(12)}(0) \sigma_{(34)}(1) \sigma_{(12)}(u) \sigma_{(34)}(\infty) \rangle \nonumber\\
&\qquad\qquad\qquad=\langle \sigma_{(12)}(0) \sigma_{(12)}(u)\rangle\langle \sigma_{(34)}(1) \sigma_{(34)}(\infty)\rangle =|u|^{-2h_2^0}\ , \\
&\langle \sigma_{(12)}(0) \sigma_{(34)}(1) \sigma_{(34)}(u) \sigma_{(12)}(\infty) \rangle \nonumber\\
&\qquad\qquad\qquad=\langle \sigma_{(12)}(0) \sigma_{(12)}(\infty)\rangle\langle \sigma_{(34)}(1) \sigma_{(34)}(u)\rangle =|1-u|^{-2h_2^0}\ ,
\end{align}
\end{subequations}
where $h_2^0$ is given by \eqref{eq:conformal-weight-single-cycle}. Reinstating the combinatorical factors discussed in Section~\ref{sec:symm orbifold}, we hence have
\begin{align} 
\langle \sigma_2(0)\sigma_2(1)\sigma_2(u)\sigma_2(\infty)\rangle_{\text{disc}}&=\frac{(2(N-2)!)^2}{N!(N-4)!}\left(1+|u|^{-\frac{c}{8}}+|1-u|^{-\frac{c}{8}}\right)\\
&=\left(4-\frac{16}{N}+\frac{8}{N^2}+\cdots\right)\left(1+|u|^{-\frac{c}{8}}+|1-u|^{-\frac{c}{8}}\right)\ .
\end{align}
This contributes at order $N^0$ to the correlator. This is expected from the counting \eqref{eq:large N behaviour}, since the covering space is the disjoint union of two spheres with two punctures each. Since the disconnected contribution to \eqref{eq:gauge-dep-corr-2222} was in some sense trivial, we focus now on the connected piece, and specifically the genus 0 contribution.
\subsection{The covering map} \label{subsec:2222 covering map}
As a first step, we work out the covering map in this case following the recipe of Section~\ref{subsec:parametrisation covering map}. In this instance, there is just one $\ell$, satisfying the equation
\be 
\frac{1}{\ell}+\frac{1}{\ell-1}+\frac{1}{\ell-x}=0\ ,
\ee
which has the solution
\be 
\ell=\frac{1}{3}\left(1+x\pm \sqrt{1-x+x^2}\right)\ .
\ee
There are two solutions, which will correspond to the two different possible OPE channels. The derivative of the covering map takes the form
\be 
\partial \Gamma(t)=\mathcal{N} \frac{t(t-1)(t-x)}{(t-\ell)^2}\ .
\ee
It is convenient to express quantities in terms of $\ell$, since in this variable they are all single-valued.
One calculates
\begin{align}
\mathcal{N}&=4\ell-2\ , \\
\Gamma(t)&=\frac{t^2 (2 \ell t-3 \ell-t+2)}{t-\ell}\ .
\end{align}
It then also follows
\begin{subequations}
\begin{align}
a_1&=\frac{3 \ell-2}{\ell}\ , \\
a_2&=-\frac{3 \ell-1}{\ell-1}\ , \\
a_3&=\frac{(2 \ell-1) (3 \ell-2) (3 \ell-1)}{(\ell-1) \ell}\ , \\
a_4&=2\ell-1\ , \\
C&=2 (\ell-1)^2 \ell^2\ .
\end{align}
\end{subequations}
Both the cross-ratio in the covering space, as well as the cross-ratio in the base space are expressible in terms of $\ell$ and take the form
\be 
u=\frac{\ell (3 \ell-2)^3}{2 \ell-1}\ , \qquad x=\frac{\ell (3 \ell-2)}{2 \ell-1}\ .
\ee
We should note that given $u$, there are generically four distinct solutions for $\ell$ and hence for $x$. One can also say this by noting that $x$ and $u$ satisfy the polynomial
\be 
x^4-4 u x^3+6 u x^2-4 u x+u^2=0\ , \label{eq:xu polynomial}
\ee
which has four solutions for $x$ for a given $u$.\footnote{The discriminant of this polynomial is $\text{Disc}(u)=-6912u^4(u-1)^4$,
so this conclusion is true as long as the configuration is not degenerate.} These four solutions specify four distinct covering spaces of the base sphere. This number is the Hurwitz number in this instance, and matches with what we expected from \eqref{eq:4pt Hurwitz number}.

The polynomial \eqref{eq:xu polynomial} is irreducible, which means that all roots $x=x(u)$ are related by analytic continuation in $u$. 
This is the reflection of our claim at the end of Section~\ref{subsec:large N} that there is only one correlation function to compute, the other three are obtained by analytic continuation.
\subsection{The four-point function}
We now compute the sphere contribution to the connected four-point function which according to \eqref{eq:final result four point function} takes the form
\be 
\langle \sigma_{g_1}(0)\sigma_{g_2}(1)\sigma_{g_3}(u)\sigma_{g_4}(\infty)\rangle =\left|256(\ell-1)^3\ell^3(3\ell-2)(3\ell-1)\right|^{-\frac{c}{12}}\ ,\label{eq:four point function 2222}
\ee
where the specification of the group representatives picks out one specific solution $\ell=\ell(u)$.

This solves the differential equation \eqref{eq:diff-eq-ssss-3}, which in this case takes the form (when expressed in terms of $\ell$)
\be 
\left(\partial_\ell+\frac{c (2 \ell-1) \left(6 \ell^2-6 \ell+1\right)}{4(\ell-1) \ell (3 \ell-2) (3 \ell-1)}\right)\langle \sigma_{g_1}(0)\sigma_{g_2}(1)\sigma_{g_3}(u)\sigma_{g_4}(\infty)\rangle=0\ .
\ee
To determine the full gauge invariant correlator, we have to reinstate combinatorical factors and sum over the four different solutions of $\ell=\ell(u)$. This yields
\begin{multline}
\langle \sigma_2(0)\sigma_2(1)\sigma_2(u)\sigma_2(\infty)\rangle_\text{c}=\frac{(2(N-2)!)^2}{N!(N-3)!} \times \\
\times \sum_{\genfrac{}{}{0pt}{}{\ell}{\frac{\ell(3\ell-2)^3}{2\ell-1}=u}}\!\!\!\!\!\!\! \left|256(\ell-1)^3\ell^3(3\ell-2)(3\ell-1)\right|^{-\frac{c}{12}}\!\! . \label{eq:four point function 2222 gauge independent}
\end{multline}
The prefactor has expansion
\be 
\frac{(2(N-2)!)^2}{N!(N-3)!}=\frac{4}{N}-\frac{4}{N^2}+\cdots
\ee
and hence contributes at order $N^{-1}$. This is the expected behaviour from a four-punctured sphere, see eq.~\eqref{eq:large N behaviour}.

\subsection{Factorisation}
Next, let us demonstrate explicitly the factorisation in this case.\footnote{A similar analysis has been performed in Appendix B of \cite{Pakman:2009zz} and investigated more in general in \cite{Burrington:2018upk}.} We do this again for the gauge dependent correlator \eqref{eq:four point function 2222}. Let us look at the limit $x \to 0$ (or equivalently $u \to 0$). In terms of $\ell$, this either implies $\ell\to \tfrac{2}{3}$ or $\ell \to 0$. These are exactly the two possible behaviours we discussed in Section~\ref{subsec:limit x0} and they correspond to $k_1=0$ and $k_1=1$ respectively. Thus, we expect that they correspond to the two different OPE channels. Indeed, we find
\begin{enumerate}
\item $\ell \to 0$. In this limit, we have
\be 
\langle \sigma_{g_1}(0)\sigma_{g_2}(1)\sigma_{g_3}(u)\sigma_{g_4}(\infty)\rangle\sim \frac{1}{|u|^{\frac{c}{4}}}+\cdots\ . \label{eq:OPE-l=0-2222}
\ee
This is indeed the behaviour as expected in the identity channel, as
\be 
\frac{c}{4} = 2 (h_2^0+ h_2^0 - h_1^0)\ ,
\ee
where $h_w^0$ is given by \eqref{eq:conformal-weight-single-cycle}.
\item $\ell\to \tfrac{2}{3}$. One computes directly that
\be 
\langle \sigma_{g_1}(0)\sigma_{g_2}(1)\sigma_{g_3}(u)\sigma_{g_4}(\infty)\rangle\sim \frac{2^{-\frac{8c}{9}}3^{\frac{c}{2}}}{|u|^{\frac{c}{36}}}+\cdots\ . \label{eq:OPE-l=2/3-2222}
\ee
We expect this to correspond to the channel in which the twist 3 field propagates and indeed we have
\be 
\frac{c}{36} = 2 (h_2^0+h_2^0-h_3^0)  \ . 
\ee
\end{enumerate}
Note that in \eqref{eq:four point function 2222 gauge independent} we sum over different solutions of $\ell$. Hence, both the fusion channels $2 \times 2 \to 1$ and $2 \times 2 \to 3$ can be observed in the final gauge independent correlator.

\section{More differential equations for correlators}
\label{sec:more-differential-equations}
The results of the previous sections can be extended to compute more general correlators. In this section, we relax the condition for the fields in the correlator to be bare twist-fields. We first generalise the previous analysis to the case where only one of the fields is the bare twist field. We then discuss how the computation can be adapted if the null vector occurs at level 2. Finally, we study multi-cycle and higher-point correlators of bare twist fields. 

\subsection{One twist field}\label{subsec:one twist field}
As we already noticed in Section \ref{sec:null-vectors-into-differential-equations}, we only made use of the null vector $L_{-1/w_3}\sigma_{g_3}=0$. It follows that eq.~\eqref{eq:diff-eq-ssss-1} can be easily generalized to 
\be 
\left(\alpha_3 \partial_u+(h_1\gamma_1+h_2\gamma_2+h_3^0 \gamma_3-h_4\gamma_4)\right)\left\langle \mathcal{O}_{g_1}(0) \mathcal{O}_{g_2}(1) \sigma_{g_3}(u) \mathcal{O}_{g_4}(\infty) \right\rangle=0\ .\label{eq:diff-eq-VVsV-1}
\ee
where $\mathcal{O}_{g_j}$ for $j = 1$, $2$, $4$ is a primary field of conformal weight $h_j$ in the $w_j$-twisted sector. 

The functions $f_k(z)$ can be chosen as in eqs.~\eqref{eq:f_k=h-comp-Gamma-minus1} and \eqref{eq:h=dixGamma}. We find the differential equation
\begin{multline}
\left[ \partial_x +  h_1 \frac{\partial_x a_1}{a_1} + h_2 \frac{\partial_x a_2}{a_2} + \frac{c}{24} \frac{w_3^2-1}{w_3} \frac{\partial_x a_3}{a_3} - h_4  \frac{\partial_x a_4}{a_4} - \frac{c}{12} \left(\frac{w_3^2-1}{w_3} \right) \frac{b_3}{a_3} \right] \times \\
\times \left\langle \mathcal{O}_{g_1}(0) \mathcal{O}_{g_2}(1) \sigma_{g_3}(u) \mathcal{O}_{g_4}(\infty) \right\rangle = 0 \ .
\end{multline}
Making use of identity \eqref{eq:abC relation}, this becomes 
\begin{multline}
\Biggl[ \partial_x + \left(h_1 - \frac{c}{24}(w_1-1) \right) \frac{\partial_x a_1}{a_1} + \left(h_2 - \frac{c}{24}(w_2-1) \right) \frac{\partial_x a_2}{a_2} +\dfrac{c}{24}  \left(\frac{w_3-1}{w_3} \right) \frac{\partial_x a_3}{a_3}\\
   - \left(h_4 - \frac{c}{24}(w_4-1) \right) \frac{\partial_x a_4}{a_4} + \frac{c}{12} \sum_{i} \frac{\partial_x C_i}{C_i} \Biggr]\left\langle \mathcal{O}_{g_1}(0) \mathcal{O}_{g_2}(1) \sigma_{g_3}(u) \mathcal{O}_{g_4}(\infty) \right\rangle = 0 \  ,
\end{multline}
and taking into account holomorphic and antiholomorphic contributions we obtain
\begin{multline} 
\left\langle \mathcal{O}_{g_1}(0) \mathcal{O}_{g_2}(1) \sigma_{g_3}(u) \mathcal{O}_{g_4}(\infty) \right\rangle = \text{const.} \ |a_1|^{-2h_{1} + \frac{c}{12}(w_1-1)} |a_2|^{-2 h_{2} + \frac{c}{12}(w_2-1)}  \\
\times |a_3|^{-\frac{c}{12}\frac{w_3-1}{w_3}} |a_4|^{2 h_{4} - \frac{c}{12}(w_4-1)} \left(\prod_i |C_i| \right)^{-\frac{c}{6}} \  , 
\label{eq:H(x)}
\end{multline}
where the integration constant is left undetermined. The normalisation of \eqref{eq:H(x)} depends now also on the specific seed theory we are taking the orbifold of. 

\subsection{A twisted BPZ equation}

In the derivation of eq.~\eqref{eq:H(x)}, as well as in the one of \eqref{eq:solution diff eq}, our method relied on the existence of the null vector $L_{-1/w} \sigma_{g_w}$. One can wonder if it is possible to derive differential equations for correlators by exploiting the existence of higher level null vectors. We find that this is actually possible. To illustrate this, we consider the case where the seed theory $\mathcal{M}$ has only Virasoro symmetry. We will show how to derive and solve a differential equation for the correlator
\be 
\left\langle \mathcal{O}_{g_1}(0) \mathcal{O}_{g_2}(1) \mu_{g_3}(u) \mathcal{O}_{g_4}(\infty) \right\rangle \  ,
\label{eq:level2-correlator}
\ee
where $\mu_{g_3}$ has a null vector at level 2.
We shall adopt a convenient way to parametrise the conformal weight as follows,
\be 
h_j = h_j^0+\frac{\Delta_j}{w_j} \  .
\ee
This is motivated by the fact that $\Delta_j$ can be interpreted as the conformal weight in the covering space in the covering space method (see \eqref{eq:Virasoro-algebra-isomorphism}). Moreover, as is customary for a Virasoro symmetry, we parametrise the central charge as $c = 1 + 6(b+b^{-1})^2$. 
The different fields are assumed to be defined in the twisted sectors $w_j$ and we leave $h_j$ (or equivalently $\Delta_j$) for $j \ne 3$ generic.
The `covering space conformal weight' of $\mu_{g_3}(z)$ is fixed by the existence of the null vector at level 2 and reads
\be 
\Delta_{3} = - \left(\frac{1}{2} + \frac{3}{4}b^{-2} \right)  \ .
\label{eq:h-mu}
\ee 
For this value of the conformal weight, the field 
\be 
\left( L_{-\frac{2}{w_3}} + w_3 b^{2} L_{-\frac{1}{w_3}}^2 \right) \mu_{g_3} 
\label{eq:level2-null-vector}
\ee
is null and hence decouples from correlation functions
\be 
\left\langle \mathcal{O}_{g_1}(0) \mathcal{O}_{g_2}(1) \left(\left( L_{-\frac{2}{w_3}} + w_3 b^{2} L_{-\frac{1}{w_3}}^2 \right)\mu_{g_3}\right)(u) \mathcal{O}_{g_4}(\infty) \right\rangle = 0 \  . 
\label{eq:null-rel-VVpiV}
\ee
We are going to show how a differential equation for the correlator \eqref{eq:level2-correlator} can be deduced from this null-vector constraint. 

Let us start by noticing that repeating the same arguments made in Section~\ref{sec:diffeq-ssss} to deduce eq.~\eqref{eq:Fsigmas=0}, we find
\be 
\oint_C \mathrm{d}z \left\langle \sum_{k = 1}^n f_{k}(z) T^{(k)}(z) \mathcal{O}_{g_1}(0) \mathcal{O}_{g_2}(1) \mu_{g_3}(u) \mathcal{O}_{g_4}(\infty) \right\rangle = 0 \  .\label{eq:level2 null contour}
\ee
It follows
\begin{multline}
\left(\alpha_3 \partial_u+(h_1\gamma_1+h_2\gamma_2+h_3\gamma_3-h_4\gamma_4)\right)\left\langle \mathcal{O}_{g_1}(0) \mathcal{O}_{g_2}(1) \mu_{g_3}(u) \mathcal{O}_{g_4}(\infty) \right\rangle \\ 
+ \beta_3 \left\langle \mathcal{O}_{g_1}(0) \mathcal{O}_{g_2}(1) \left( L_{-\frac{1}{w_3}} \mu_{g_3}\right)(u) \mathcal{O}_{g_4} (\infty)  \right\rangle  =0 \  ,
\label{eq:diff-eq-VVpiV-1}
\end{multline}
where the term proportional to $\beta_3$ does not vanish in this case. Similarly, from
\be 
\oint_C \mathrm{d}z \left\langle \sum_{k = 1}^n f_{k}(z) T^{(k)}(z)\mathcal{O}_{g_1}(0) \mathcal{O}_{g_2}(1) \left( L_{-\frac{1}{w_3}} \mu_{g_3}\right)(u) \mathcal{O}_{g_4}(\infty) \right\rangle = 0 
\ee
we obtain 
\begin{multline}
\left( \alpha_3 \partial_u +  \left(h_3 + \tfrac{1}{w_3}\right) \gamma_3 + h_1 \gamma_1 + h_2 \gamma_2 - h_4 \gamma_4   \right) \left\langle \mathcal{O}_{g_1}(0)\mathcal{O}_{g_2}(1) \left( L_{-\frac{1}{w_3}} \mu_{g_3}\right)(u) \mathcal{O}_{g_4}(\infty)  \right\rangle \\
+ \beta_3 \left\langle \mathcal{O}_{g_1}(0)\mathcal{O}_{g_2}(1) \left( L_{-\frac{1}{w_3}}^2 \mu_{g_3}\right)(u) \mathcal{O}_{g_4}(\infty)  \right\rangle \\
+ \delta_3 \left( \frac{2}{w_3}h_3 -\frac{c}{12}\left( \frac{w_3^2-1}{w_3^2}\right) \right)\left\langle \mathcal{O}_{g_1}(0)\mathcal{O}_{g_2}(1) \mu_{g_3} (u) \mathcal{O}_{g_4}(\infty) \right\rangle = 0  \  , 
\label{eq:diff-eq-VVpiV-2}
\end{multline}
where the term proportional to $\delta_3$ is due to the contribution of 
\be 
\left\langle \mathcal{O}_{g_1}(0)\mathcal{O}_{g_2}(1) \left( L_{\frac{1}{w_3}} L_{-\frac{1}{w_3}} \mu_{g_3}\right)(u) \mathcal{O}_{g_4}(\infty)  \right\rangle \  . 
\ee
Putting together eqs.~\eqref{eq:diff-eq-VVpiV-1} and \eqref{eq:diff-eq-VVpiV-2} we find 
\begin{align}
& \beta_3^2 \left\langle \mathcal{O}_{g_1}(0)\mathcal{O}_{g_2}(1) \left( L_{-\frac{1}{w_3}}^2 \mu_{g_3}\right)(u) \mathcal{O}_{g_4}(\infty)  \right\rangle  \nonumber \\
& \quad  = \Biggl[ \left( \alpha_3 \partial_u + \left(h_3 + \tfrac{1}{w_3}\right) \gamma_3 + h_1 \gamma_1 + h_2 \gamma_2 - h_4 \gamma_4   \right) \left( \alpha_3 \partial_u + h_3 \gamma_3 + h_1 \gamma_1 + h_2 \gamma_2 - h_4 \gamma_4   \right)\nonumber\\ 
& \qquad\qquad\qquad\quad - \frac{2\beta_3 \delta_3 \Delta_3}{w_3} \biggr] \left\langle \mathcal{O}_{g_1}(0)\mathcal{O}_{g_2}(1) \mu_{g_3} (u) \mathcal{O}_{g_4}(\infty) \right\rangle \ .  \label{eq:diff-eq-VVpiV-3} 
\end{align}
In order to express in terms of differential operators the term containing $L_{-\frac{2}{w_3}}$ in eq.~\eqref{eq:null-rel-VVpiV}, we use a relation similar to \eqref{eq:level2 null contour}, but with different functions $f_k$, which we denote by $\tilde f_k(z)$. They obey the following properties: 
\begin{enumerate}
\item Close to the insertion points $z_j$, for all $k \in g_j$ \label{item 1tilde}
\be 
\tilde f_{k}(e^{2 \pi i}z +z_j) = \tilde f_{g_j(k)}(z+z_j) \  , \qquad j = 1,\dots,4 \  .  
\label{eq:tilde f-monodromy}
\ee 
This ensures single-valuedness of the integrand.
\item $\tilde f_k(z)$ has no poles except at infinity, where the behaviour is specified through item~\ref{item 3tilde}.\label{item 2tilde}
\item Close to the insertion points $z_j$, for all $k \in g_j$, \label{item 3tilde}
\begin{subequations} \label{eq:tildefexp}
\begin{align}
\tilde f_k(z) &=  \tilde \gamma_j (z-z_j) + \mathcal{O}\left((z-z_j)^{\frac{w_j+1}{w_j}}\right) \ ,  && z \sim z_j\,,\ j=1,\,2 \ , \label{eq:tildefexp1} \\
\tilde f_k(z) & \sim \tilde \alpha_3 + \tilde \eta_3 \mathrm{e}^{-\frac{4 \pi i k}{w_3}} (z-u)^{\frac{w_3-2}{w_3}} + \tilde \beta_3  \mathrm{e}^{-\frac{2\pi i k}{w_3}} (z-u)^{\frac{w_3-1}{w_3}}\hspace{-30pt} \nonumber \\
& \hspace{50pt} + \tilde \gamma_3(z-u) + \mathcal{O}\left((z-u)^{\frac{w_3+1}{w_3}}\right)\ , && z \sim u \ ,   \label{eq:tildefexp2} \\
\tilde f_k(z) &=   \tilde \gamma_4 z+ \mathcal{O}\left(z^{\frac{w_4-1}{w_4}}\right) \  , &&z \sim \infty\ .  \label{eq:tildefexp3}
\end{align}
\end{subequations}
where the $w_r$ values of $k$ correspond to the different branches.  
\end{enumerate}
Thus, the only difference compared to $f_k$ is the existence of the coefficient $\tilde{\eta}_3$, which will be responsible in picking out the mode $L_{-2/w_3}$ acting on $\mu_{g_3}$.
The relation
\be 
\oint_C \mathrm{d}z \left\langle \sum_{k = 1}^n \tilde f_{k}(z) T^{(k)}(z) \mathcal{O}_{g_1}(0)\mathcal{O}_{g_2}(1) \mu_{g_3} (u) \mathcal{O}_{g_4}(\infty) \right\rangle = 0 
\ee
now leads to
\begin{multline}
\tilde \eta_3 \left\langle \mathcal{O}_{g_1}(0)\mathcal{O}_{g_2}(1) \left( L_{-\frac{2}{w_3}} \mu_{g_3}\right)(u) \mathcal{O}_{g_4}(\infty) \right\rangle \\
+  \tilde \beta_3 \left\langle \mathcal{O}_{g_1}(0)\mathcal{O}_{g_2}(1) \left( L_{-\frac{1}{w_3}} \mu_{g_3}\right)(u) \mathcal{O}_{g_4}(\infty)  \right\rangle \\
+(\tilde \alpha_3 \partial_u + h_1 \tilde \gamma_1 + h_2 \tilde \gamma_2 + h_3 \tilde \gamma_3 - h_4 \tilde \gamma_4)\left\langle \mathcal{O}_{g_1}(0)\mathcal{O}_{g_2}(1) \mu_{g_3} (u) \mathcal{O}_{g_4}(\infty) \right\rangle = 0 \ . 
\label{eq:diff-eq-VVpiV-4}
\end{multline}
Combining \eqref{eq:null-rel-VVpiV}, \eqref{eq:diff-eq-VVpiV-1}, \eqref{eq:diff-eq-VVpiV-3} and \eqref{eq:diff-eq-VVpiV-4} we find
\begin{align} 
\Biggl[ & -\beta_3^2 (\tilde \alpha_3 \partial_u + h_1 \tilde \gamma_1 + h_2 \tilde \gamma_2 + h_3 \tilde \gamma_3 - h_4 \tilde \gamma_4)  + \beta_3 \hspace{1pt} \tilde \beta_3 \left( \alpha_3 \partial_u +  h_3 \gamma_3 + h_1 \gamma_1 + h_2 \gamma_2 - h_4 \gamma_4   \right) \nonumber \\
 & \quad  + w_3 \hspace{1pt} b^2 \hspace{1pt} \tilde \eta_3 \left( \alpha_3 \partial_u + \left(h_3 + \tfrac{1}{w_3}\right)\gamma_3 + h_1 \gamma_1 + h_2 \gamma_2 - h_4 \gamma_4   \right)  \nonumber\\
& \hspace{140pt} \times \left( \alpha_3 \partial_u + h_3 \gamma_3 + h_1 \gamma_1 + h_2 \gamma_2 - h_4 \gamma_4   \right) \nonumber \\
& \hspace{130pt} - 2 \hspace{1pt} b^2 \hspace{1pt} \tilde \eta_3 \hspace{1pt} \beta_3 \hspace{1pt} \delta_3 \hspace{1pt} \Delta_3 \Biggr]\left\langle \mathcal{O}_{g_1}(0)\mathcal{O}_{g_2}(1) \mu_{g_3} (u) \mathcal{O}_{g_4}(\infty) \right\rangle = 0 \  . \label{eq:level2-diff-eq}
\end{align}

\subsubsection[Constructing the functions \texorpdfstring{$\tilde f_k$}{fk}]{Constructing the functions $\boldsymbol{\tilde f_k}$}

The functions $f_k(z)$ can once more be chosen according to eqs.~\eqref{eq:f_k=h-comp-Gamma-minus1} and \eqref{eq:h=dixGamma}. 
Regarding the functions $\tilde f_k(z)$, item~\ref{item 1tilde} is satisfied by the ansatz 
\begin{equation}
\tilde f_k = \tilde h \circ \Gamma_k^{-1} \  ,
\end{equation}
where $\tilde h$ is a meromorphic function on the covering space. One can show that items~\ref{item 2tilde} and \ref{item 3tilde} can be satisfied by setting 
\be 
\tilde h(t) = \frac{t(t-1)}{t-x} \partial_t \Gamma(t) \  . 
\ee
The coefficients entering \eqref{eq:level2-diff-eq} can be expressed in terms of the covering map \eqref{eq:Gamma expansions} as follows,
\begin{subequations} \label{eq:fkt parameters}
\begin{align}
\tilde \alpha_3 & = 0  \  ,  \\
\tilde \eta_3 & = w_3 \hspace{1pt} x \hspace{1pt} (x-1) \hspace{1pt} a_3^{\frac{2}{w_3}} \  , \\
\tilde \beta_3 & = a_3^{-1 + \frac{1}{w_3}} (-3 x \hspace{1pt} b_3 + 3 x^2 \hspace{1pt} b_3 - w_3 \hspace{1pt} a_3 + 2 x \hspace{1pt} w_3 \hspace{1pt} a_3 )  \  ,  \\
\tilde \gamma_1 & = \frac{w_1}{x} \  , \\
\tilde \gamma_2 & = \frac{w_2}{1-x} \  , \\
\tilde \gamma_3 & =  w_3 - \frac{x(x-1)(2 w_3-1) \hspace{1pt} b_3^2}{w_3 a_3^2} + \frac{(4x-2) b_3 + 4 x (x-1) c_3}{a_3} \  , \\
\tilde \gamma_4 & = w_4  \  .\
\end{align}
\end{subequations}
For the coefficients entering the expansion of $f_k(z)$, in addition to eq.~\eqref{eq:fk parameters}, we have
\begin{subequations} \label{eq:beta3-delta3}
\begin{align}
\beta_3 & = -w_3 a_3^{\frac{1}{w_3}}  \  ,  \\
 \delta_3 & = \frac{a_3^{-2 - \frac{1}{w3}}}{2 w_3} \left(-2 w_3 \hspace{1pt} b_3 \hspace{1pt} \partial_x a_3 - 6 w_3 \hspace{1pt} a_3 \hspace{1pt} c_3 + 2 w_3 \hspace{1pt} a_3 \hspace{1pt} \partial_x b_3 + (3 w_3+1) b_3^2 \right) \  . 
\end{align}
\end{subequations}

\subsubsection{Solving the differential equation}

In order to simplify eq.~\eqref{eq:level2-diff-eq}, it is convenient to make the ansatz
\begin{multline}
\left\langle \mathcal{O}_{g_1}(0)\mathcal{O}_{g_2}(1) \mu_{g_3} (u) \mathcal{O}_{g_4}(\infty) \right\rangle = |a_1|^{-2h_{1} + \frac{c}{12}(w_1-1)} |a_2|^{-2 h_{2} + \frac{c}{12}(w_2-1)} |a_3|^{-2 h_{3} + \frac{c}{12}(w_3-1)} \\
\times  |a_4|^{2 h_{4} - \frac{c}{12}(w_4-1)} \left(\prod_i |C_i| \right)^{-\frac{c}{6}} M(x,\bar x)
\label{eq:ansatz-level2}
\end{multline}
and solve for $M(x,\bar x)$. Here, we stripped off the conformal factors that are expected from the analysis of \cite{Lunin:2000yv, Pakman:2009zz} and $M(x,\bar{x})$ is the correlation function in the covering space. Compare this to eq.~\eqref{eq:H(x)}.
Making use of the identities \eqref{eq:abC relation}, \eqref{eq:abC relation-bis} and of eqs.~\eqref{eq:fk parameters}, \eqref{eq:fkt parameters}, \eqref{eq:beta3-delta3}, the differential equation \eqref{eq:level2-diff-eq} collapses to the BPZ \cite{Belavin:1984vu} equation for $M(x,\bar{x})$, 
\be  
\left( \frac{x (1-x)}{b^2} \frac{\text{d}^2}{\text{d}x^2} + (2x-1) \frac{\text{d}}{\text{d}x} - \frac{3 b^2+2}{4} +\frac{\Delta_1}{x}+\frac{\Delta_2}{1-x}-\Delta_4 \right)M(x,\bar{x}) = 0 \  .
\label{eq:BPZ}
\ee
This differential equation can be solved in terms of hypergeometric functions, see e.g.~\cite{Ribault:2014hia} for a detailed discussion. This result hence makes equation \eqref{eq:relation base space covering space} precise, since the final correlator indeed has a universal prefactor given by covering map data, which we factored out in \eqref{eq:ansatz-level2}, times the correlator evaluated on the covering space.

\subsection{Higher-point and multi-cycle correlation functions}
\label{sec:higher-point-and-multi-cycle-corr}

In this section we will show that the method described in Section~\ref{sec:diffeq-ssss} to compute four-point functions of twist operators can be directly extended to higher-point functions. We find that the answer has exactly the same structure as in the four-point function case. In particular, as already alluded to in Section~\ref{subsec:large N}, the coincidence limit of higher-point functions also yields correlators of multi-cycle twist fields.
This is because e.g.
\be 
\sigma_{(1 \cdots w_1)(w_1+1 \cdots w_1+w_2)}(z_2)\sim \sigma_{(1 \cdots w_1)}(z_1) \sigma_{(w_1+1\cdots w_1+w_2)}(z_2) \ , \qquad z_1 \to z_2 \ , 
\ee
see eq.~\eqref{eq:sigma-sigma-OPE}. As a consequence the answer for the sphere contribution to the twist correlators is completely universal and is valid for arbitrary higher-point functions and for multi-cycle twist fields.

For simplicity, let us illustrate our method by deriving differential equations for the five-point function
\be 
G(u_1, u_2) = \left\langle \sigma_{g_1}(0) \sigma_{g_2}(1) \sigma_{g_3}(u_1) \sigma_{g_4}(u_2) \sigma_{g_5}(\infty) \right\rangle \ . 
\label{eq:five-pt-fct}
\ee
The derivation for higher-point functions will be analogous. Note that the correlation function \eqref{eq:five-pt-fct} depends now on two cross-ratios, $u_1$ and $u_2$. The arguments of Section~\ref{sec:diffeq-ssss} can be repeated and an essentially identical derivation implies in the present context
\be 
\left( \alpha_3^a \partial_{u_1} + \alpha_4^a \partial_{u_2} + h^0_1 \gamma_1^a + h^0_2 \gamma_2^a + h^0_3 \gamma_3^a + h^0_4 \gamma_4^a - h^0_5 \gamma_5^a \right) G(u_1, u_2) = 0 \ , 
\label{eq:five-s-diff-eq}
\ee
where $\alpha_3^a$, $\alpha_4^a$ and $\gamma_1^a , \dots, \gamma_5^a$ enter the expansion of a function $f^a_k(z)$ satisfying the following properties: 
\begin{enumerate}
\item Close to the insertion points $z_j$, for $k = 1, \dots, n$  \label{item 1a}
\be 
f_{k}^a(e^{2 \pi i}z +z_j) = f^a_{g_j(k)}(z+z_j) \  , \qquad j = 1,\dots,5 \  .  
\label{eq:fa-monodromy}
\ee 
\item $f_k^a(z)$ has no poles except at infinity, where the behaviour is specified through item~\ref{item 3}.\label{item 2a}
\item Close to the insertion points $z_j$, for $k = 1, \dots, n$, the functions $f_k^a$ have expansions \label{item 3a}
\begin{subequations} \label{eq:fa-expansion}
\begin{align} 
f_k^a(z) & \sim  \gamma_1^a z + \dots \ , \qquad && z \sim 0 \ ,  \\
f_k^a(z) & \sim  \gamma_2^a (z-1) + \dots \ , \qquad && z \sim 1 \ ,  \\
f_k^a(z) & \sim \alpha_3^a + \beta_3^a \mathrm{e}^{-\frac{2\pi i k}{w_3}} (z-u_1)^{\frac{w_3-1}{w_3}} + \gamma_3^a(z-u_1) +  \dots \ , \qquad && z \sim u_1 \ ,  \\
f_k^a(z) & \sim \alpha_4^a + \beta_4^a \mathrm{e}^{-\frac{2\pi i k}{w_4}} (z-u_2)^{\frac{w_4-1}{w_4}} + \gamma_4^a(z-u_2) +  \dots \ , \qquad && z \sim u_2 \ ,  \\
f_k^a(z) & \sim  \gamma_5^a z+ \dots \qquad && z \sim \infty \ . 
\end{align}
\end{subequations}
\end{enumerate}
Note that the auxiliary functions $f_k^a$ carry an extra label $a\in \{1,2\}$. This is because there are now \emph{two} independent functions obeying the requirements in items \ref{item 1a}--\ref{item 3a}, namely
\be 
f^1_k = \partial_{x_1} \Gamma \circ \Gamma_{k}^{-1} \ , \qquad f^2_k = \partial_{x_2} \Gamma \circ \Gamma_{k}^{-1} \ , 
\ee
where \eqref{eq:Gamma expansions} is now replaced by 
\begin{subequations}\label{eq:Gamma expansions five}
\begin{align}
\Gamma(t)\sim & \ 0 + a_1 t^{w_1}+b_1 t^{w_1+1}+\mathcal{O}(t^{w_1+2}) & \hspace{5pt} t &\sim 0\ , \label{eq:Gamma-at-0-five}\\
\Gamma(t)\sim & \ 1 + a_2(t-1)^{w_2}+b_2(t-1)^{w_2+1}+\mathcal{O}((t-1)^{w_2+2}) & t &\sim 1\ , \\
\Gamma(t)\sim & \ u_1 + a_3(t-x_1)^{w_3}+b_3(t-x_1)^{w_3+1}+ \mathcal{O}((t-x_1)^{w_3+2})\!\!\! & t &\sim x_1\ , \\
\Gamma(t)\sim & \ u_2 + a_4(t-x_2)^{w_4}+b_3(t-x_2)^{w_4+1}+ \mathcal{O}((t-x_2)^{w_4+2})\!\!\! & t &\sim x_2\ , \\
\Gamma(t)\sim & \ a_5 t^{w_5} + b_5 t^{w_5-1} + \mathcal{O}(t^{w_5-2}) & t &\sim \infty\ .  \label{eq:Gamma-at-inf-five}
\end{align}
\end{subequations}
Note that $G(u_1, u_2)$ obeys \emph{two} differential equations, i.e.~eq.~\eqref{eq:five-s-diff-eq} for $a=1$, $2$. By rewriting the parameters entering eq.~\eqref{eq:fa-expansion} in terms of covering map data and making use of identity \eqref{eq:abC relation-many}, the differential equations in \eqref{eq:five-s-diff-eq} become for $a=1$, $2$, 
\begin{multline}
\Biggl[ \partial_{x_a} + \dfrac{c}{24} \sum_{j=1}^4 \frac{(w_j-1)}{w_j} \frac{\partial_{x_a} a_j}{a_j} - \frac{c}{24} \frac{(w_5-1)}{w_5} \frac{\partial_{x_a} a_5}{a_5} + \frac{c}{12} \sum_{i}\frac{\partial_{x_a} C_i}{C_i} \Biggr] G(x_1,x_2) = 0 \  , 
\label{eq:final-diff-eq-five-pt}
\end{multline}
where $x_1$ and $x_2$ are the two cross-ratios in the covering space and $\Gamma(x_1) = u_1\ , \Gamma(x_2) = u_2$. Eq.~\eqref{eq:final-diff-eq-five-pt} can be easily solved and putting together the contribution from left and right movers, we find 
\be 
G_{g_1,\dots,g_5}(x_1, x_2, \bar{x}_1, \bar{x}_2) = \left| C_{w_1, \dots, w_5} \right|^2  \prod_{j=1}^4 |a_j|^{-\frac{c(w_j-1)}{12w_j}} |a_5|^{\frac{c(w_5-1)}{12w_5}} \prod_i |C_i|^{-\frac{c}{6}} \  ,
\ee
The result is analogous for an $m$-point function of bare twist operators,  
\begin{align}
& \left\langle \sigma_{g_1}(0,0) \sigma_{g_2}(1,1) \sigma_{g_3}(u_1, \bar u_1) \dots \sigma_{g_{m-1}}(u_{m-3},\bar u_{m-3}) \sigma_{g_{m}}(\infty, \infty) \right\rangle \nonumber \\
& \qquad = \left| C_{w_1, \dots, w_{m}} \right|^2  \prod_{j=1}^{m-1} |a_j|^{-\frac{c(w_j-1)}{12w_j}}   |a_{m}|^{\frac{c(w_{m}-1)}{12w_{m}}}\prod_i |C_i|^{-\frac{c}{6}} \  . 
\end{align}
The constant $C_{w_1, \dots, w_{m}}$ can again be fixed by requiring factorisation and one can show that
\be 
|C_{w_1,\dots,w_\nif}|^2 =  \prod_{j=1}^{\nif-1} w_j^{-\frac{c(w_j+1)}{12}} w_\nif^{\frac{c(w_\nif+1)}{12}}\ .
\ee

\section{Higher genus correlators}\label{sec:higher genus}
In this section, we discuss how our method can be applied if the covering surface has higher genus.

\subsection{Parameter counting}
Let us begin by analysing more structurally the constraints on the functions $f_k$, which played a crucial role in evaluating the null-vector constraint in Section~\ref{sec:diffeq-ssss}. We discuss for simplicity the four-point function case. The analysis for higher-point functions is very similar.

Deriving differential equations for correlators started from writing\footnote{We keep the insertion points general, since the following discussion is somewhat easier if no field is at $\infty$. The result is the same in either case.}
\be 
\oint_C \mathrm{d}z\ \left\langle \sum_k f_k(z)T^{(k)}(z) \mathcal{O}_{g_1}(z_1) \mathcal{O}_{g_2}(z_2) \sigma_{g_3}(z_3) \mathcal{O}_{g_4}(z_4)\right\rangle =0\ ,\label{eq:vanishing contour}
\ee
where the contour $C$ encircles all insertion points. The fields $\mathcal{O}_{g_i}$ are primary twisted fields, twisted by the group element $g_i$. We chose the functions $f_k(z)$ such that the resulting correlator becomes single-valued. This is achieved by setting 
\be 
f_k=h \circ \Gamma_k^{-1}\ ,
\ee
where $k$ is the $k$-th branch of the inverse of the covering map. $h$ is a meromorphic function on the covering space. We imposed that $h$ can have at most double poles at the poles of $\Gamma(t)$, since this leads to an asymptotic growth $f_k(z) =\mathcal{O}(z^2)$ near $z=\infty$. Thus, there is no contribution from $\infty$ in the integral \eqref{eq:vanishing contour}. We also imposed that $h$ should not have any other poles away from the poles of $\Gamma(t)$. Denoting the poles of $\Gamma(t)$ as before by ${\ell_1, \dots, \ell_n}$, $h$ is hence a meromorphic function with poles at most of order two at $\ell_i$.

We then computed the residues of the integral at the insertion points. This leads in general to all the modes\footnote{Modes $L_n$ for $n>0$ annihilate the field $\mathcal{O}_{g_j}(z_j)$ by assumption. The analysis can also be generalised to descendant fields.} 
\be 
L_{-1}\  ,\ L_{-1+\frac{1}{w_i}}\  , \ \dots\  ,\ L_{-\frac{1}{w_i}}\  ,\ \text{and}\ L_0
\ee
acting on the field $\mathcal{O}_{g_i}(z_i)$. Of these modes, we only know how to evaluate $L_{-1}$ and $L_0$ in general. Moreover, we used for the third field that there is a null vector $L_{-\frac{1}{w_3}}$, which allowed us to evaluate this mode as well. Thus, there were 
\be 
(w_1-1)+(w_2-1)+(w_3-2)+(w_4-1)=\sum_{j=1}^4 w_j-5 \label{eq:number of constraints}
\ee
unknown modes, which have to be eliminated by a suitable choice of $h$.

By the Riemann-Roch theorem, there is a $(2n+1-g)$-dimensional space of possible $h$'s satisfying the above requirements on a genus $g$ Riemann surface. The number of poles $n$ is equal to the degree of the covering map, which can be computed by the Riemann-Hurwitz formula
\be 
n=\frac{1}{2} \sum_{i=1}^4 w_i -1-g\ .
\ee
Thus, there is a $\big(\sum_{i=1}^4 w_i-1-3g\big)$-dimensional space of possible choices for $h$. There are some special choices which should not be counted. We can choose
\be 
h=1\ , \quad \Gamma\ , \quad\text{or}\quad \Gamma^2\ ,
\ee
which leads to $f_k(z)=1$, $z$ or $z^2$ and hence gives rise to the three global Ward identities.  They should be subtracted from the space of possible $h$'s (since we already incorporated their constraints into the four point function). Thus there is effectively a $\big(\sum_{i=1}^4 w_i-4-3g\big)$-dimensional space of choices of $h$'s. Comparing this with \eqref{eq:number of constraints}, we see that for $g=0$, there is generically a one-dimensional space of solutions. This is what we saw explicitly above, where the solution space was spanned by $h=\partial_x \Gamma$.

From the counting of parameters, we see that this becomes harder and harder for higher genus $g$. We will in the following discuss the next case of genus one correlators. In this case, we restrict to the case of four twist fields $\sigma_{g_i}(z_i)$. This gives us the ability to make use of more null vectors, since any of the modes $L_{-\frac{1}{w_i}}$ can be set to zero. Thus, there are only
\be 
(w_1-2)+(w_2-2)+(w_3-2)+(w_4-2)=\sum_{i=1}^4 w_i-8 \label{eq:number of constraints torus}
\ee
unknown modes in this case. The counting of $h$'s also changes slightly. Consider the choice $h=\partial_t \Gamma$. This function has a high order zero when expanding around the ramification points. Because of this, it only picks up the modes $L_{-\frac{1}{w_i}}$ and $L_0$ at the insertion points. Since the former are null by assumption, we only have $L_0$ acting. $L_0$ is just acting by multiplication, so we obtain an \emph{algebraic} equation, which does not restrict the correlator, but merely expresses an identity for the coefficients of the covering map. Thus, we should also subtract the linear span of $\partial_t \Gamma$ from our solution space in this case. Hence, there is a $\big(\sum_{i=1}^4 w_i-5-3g\big)=\big(\sum_{i=1}^4 w_i-8\big)$-dimensional  space of choices for $h$ that lead to non-trivial constraints in this case. This does generically \emph{not} suffice to eliminate all the unknown modes in \eqref{eq:number of constraints torus}.\footnote{The same conclusion can also be reached by allowing one of the twist fields to be general instead of the bare twist fields. On the covering torus, this corresponds to a one-point function and since  torus correlators still obey translation invariance, it is constant. In this case, there are $\sum_{i=1}^4 w_i-7$ modes we can evaluate. Taking $h=\partial_t \Gamma$ leads to a real constraint and thus does not have to be subtracted. Hence also for this case there will generically be no solutions.} 

According to the covering space method discussed in Section~\ref{subsec:covering space method}, where the correlator is computed by lifting it on the covering space, we did not expect this to be possible. The correlator lifted to the covering space becomes the partition function of the theory on the torus. We could not succeed in deriving a differential equation for the correlator, since this would imply a differential equation on the partition function and hence the spectrum of the theory. In general, one needs however further input to determine the spectrum of the theory, since this does not just depend on the central charge of the seed theory, but on the details of $\mathcal{M}$ itself.

We can further restrict the theory in which case there is a way out. If we assume the seed theory to be a Virasoro minimal model, we have a second \emph{independent} null vector (besides $L_{-1}\ket{0}$) in the vacuum representation. For example, the simplest minimal model is the Yang-Lee minimal model with $c=-\frac{22}{5}$, which has the null vector
\be 
\left(L_{-2}^2-\frac{3}{5}L_{-4}\right)\ket{0}=0
\ee
in its vacuum module. Hence in the twisted sector, there is the additional null field
\be 
\left(L_{-\frac{2}{w}}^2-\frac{3}{5w} L_{-\frac{4}{w}}\right)\sigma(z_1)=0\ .
\ee
This allows us to compute a further mode and lower the number of constraints \eqref{eq:number of constraints torus} by one. By parameter counting, we then expect there to be one solution to the constraints, which can turn the null vectors into differential equations. 

It should be clear that deriving differential equations for $g \ge 2$ correlators becomes hopeless (at least when following this route), since one would need even more null vectors.
\subsection[The example \texorpdfstring{$\langle \sigma_{(12)}(0)\sigma_{(12)}(1)\sigma_{(12)}(u)\sigma_{(12)}(\infty)\rangle$}{<sigma12 sigma12 sigma12 sigma12>}]{The example $\boldsymbol{\langle \sigma_{(12)}(0)\sigma_{(12)}(1)\sigma_{(12)}(u)\sigma_{(12)}(\infty)\rangle}$}

In the following, instead of working out the most general torus correlator, we will illustrate this with the simplest example
\be 
\langle \sigma| \sigma(1) \sigma(u) | \sigma \rangle\equiv \langle \sigma_{(12)}(0)\sigma_{(12)}(1)\sigma_{(12)}(u)\sigma_{(12)}(\infty)\rangle\ .
\label{eq:2222-example-genus-1}
\ee
Here and in the following, we will abbreviate $\sigma(z_i)\equiv \sigma_{(12)}(z_i)$. Moreover, it is convenient to write $| \sigma \rangle \equiv \sigma(0)$ and $\langle \sigma |\equiv \sigma(\infty)$. This correlator was already evaluated using the same methods in \cite{Dupic:2017hpb}. The following computation follows theirs closely. Despite this, it should be clear that our method is also applicable to torus four-point functions involving higher twist fields.

The correlator \eqref{eq:2222-example-genus-1} was evaluated in \cite{Lunin:2000yv} and found to be equal to
\begin{align}
\langle \sigma_{(12)}(0)\sigma_{(12)}(1)\sigma_{(12)}(u)\sigma_{(12)}(\infty)\rangle=\frac{Z(\tau,\bar{\tau})}{2^{\frac{2c}{3}} |u(1-u)|^{\frac{c}{12}}}\ ,
\label{eq:2222=partition-fct}
\end{align}
where $Z$ is the partition function of the theory and $\tau$ is the modular parameter of the covering space torus.

In view of this result, the computation is hence a way to derive a differential equation on the partition function. When rewritten in terms of the modular parameter $\tau$, this differential equation is equivalent to the well-known modular differential equations of the theory \cite{Eguchi:1986sb, Eguchi:1987qd, Mathur:1988rx, Gaberdiel:2008pr}.

The covering map $\Gamma(t)$ can be written explicitly in terms of the Weierstrass $\wp$-function:
\be 
\Gamma(t)=\frac{\wp(t;\tau)-\wp(\frac{1}{2};\tau)}{\wp(\tfrac{\tau}{2};\tau)-\wp(\frac{1}{2};\tau)}\ .
\ee
By construction,
\be 
\Gamma(\tfrac{1}{2})=0\ , \qquad \Gamma(0)=\infty\ , \qquad \Gamma(\tfrac{\tau}{2})=1\ .
\ee
Moreover, because $\wp'(\tfrac{1}{2};\tau)=\wp'(\tfrac{\tau}{2};\tau)=0$ and $\wp(t;\tau)$ has a double pole at 0, these points are ramification points of order 2.
Finally, because also $\wp'(\tfrac{\tau+1}{2};\tau)=0$, the point $\tfrac{\tau+1}{2}$ is ramified. Thus, there are four ramification points of order 2. We have
\be 
u=\Gamma(\tfrac{\tau+1}{2})=\frac{\wp(\frac{\tau+1}{2};\tau)-\wp(\frac{1}{2};\tau)}{\wp(\tfrac{\tau}{2};\tau)-\wp(\frac{1}{2};\tau)}=\frac{\vartheta_4(\tau)^4}{\vartheta_3(\tau)^4}\ .\label{eq:u tau relation}
\ee 

Next, we show how to derive a BPZ type equation for the four-point function \eqref{eq:2222-example-genus-1} from these null vectors. We show this for two examples, the Yang-Lee and the Ising minimal models. For the Yang-Lee model, this is actually almost trivial. In fact, it should be clear that one can evaluate integer-moded Virasoro modes on the fields as in the derivation of the BPZ equation \cite{DiFrancesco}.

\paragraph{Yang-Lee.} 
For the Yang-Lee minimal model, with central charge $c=-\frac{22}{5}$, we have the null correlator
\begin{align}
0&=\left \langle \sigma\left |\left(L_{-1}^2-\frac{3}{10} L_{-2}\right)\sigma(u) \sigma(1)\,  \right| \sigma \right\rangle  % \\
% &=\partial_u^2\langle \sigma\, |\, \sigma(u) \sigma(1)\, |\, \sigma \rangle-\frac{3}{10} \oint_u \frac{\mathrm{d}z}{z-u}\  \langle \sigma\, |\,  T(z)\sigma(u) \sigma(1)\, |\, \sigma \rangle\ .
\end{align}
Notice that in this case the Virasoro generators are integer-moded. Thus, this constraint can be turned into a differential equation by standard methods, which leads to
\be 
\left(\partial_u^2-\frac{3(2u-1)}{10u(1-u)}\partial_u+\frac{33}{400(1-u)^2u^2}\right)\langle\sigma\, |\, \sigma(u) \sigma(1)\,|\, \sigma \rangle=0\ .
\ee

\paragraph{Ising.}
For the Ising model, we find the null vector 
\be 
\left(L_{-1}^3 -\frac{33}{16} L_{-2}L_{-1} +\frac{93}{128} L_{-\frac{3}{2}}^2 - \frac{27}{64} L_{-3} \right ) \sigma  \ . 
\label{eq:Ising-null-vector}
\ee
Also in this case, integer modes can be turned into differential operators as in the derivation of the BPZ equation: we wrap the integration contour around the Riemann sphere and make use of global Ward identities. Hence, we only have to explain how to evaluate
\be 
\langle \sigma\, |\, (L_{-\frac{3}{2}}L_{-\frac{3}{2}}\sigma)(u) \sigma(1)\,|\, \sigma \rangle\ .
\ee
We proceed as before and choose functions $f_k$, $k=1$, $2$ with the appropriate monodromy properties. This is actually very easy in this case and we can choose
\be 
f_k(z)=\pm \frac{(z-1)^{\frac{1}{2}}z^{\frac{1}{2}}}{(z-u)^{\frac{1}{2}}}
\ee
This results from choosing the meromorphic function
\be
h(t)=\frac{\wp'(t;\tau)}{2\pi i \vartheta_3(\tau)^2(\wp(t;\tau)-\wp(\frac{\tau+1}{2};\tau))}
\ee
in the covering space.
We hence evaluate the expression
\begin{align}
&\oint_u \frac{\mathrm{d}z\, (z-1)^{\frac{1}{2}}z^{\frac{1}{2}}}{(z-u)^{\frac{1}{2}}(u-1)^{\frac{1}{2}}u^{\frac{1}{2}}}\langle \sigma\, |\, (-T^{(1)}+T^{(2)})(z)(L_{-\frac{3}{2}}\sigma)(u) \sigma(1)\,|\, \sigma \rangle \nonumber \\
&\qquad=-\frac{2u-1}{64u^3(1-u)^3}\langle\sigma\, |\,\sigma(u) \sigma(1)\,|\, \sigma \rangle-\frac{1}{4u^2(1-u)^2} \partial_u\langle\sigma\, |\,\sigma(u) \sigma(1)\,|\, \sigma \rangle\nonumber\\
&\qquad\qquad-\frac{2u-1}{2u(1-u)}\langle\sigma\, |\,(L_{-2}\sigma)(u) \sigma(1)\,|\, \sigma \rangle+\langle\sigma\, |\, (L_{-\frac{3}{2}}L_{-\frac{3}{2}}\sigma)(u) \sigma(1)\,|\, \sigma \rangle\ .
\end{align}
On the other hand, we can evaluate the same expression by wrapping the contour around the Riemann sphere. The function in the integrand multiplying the four-point function was chosen such that the integrand is single-valued and there is no contribution coming from $\langle \sigma |$. From the remaining two fields at $0$ and $1$, there is in fact also no contribution, because $L_{-\frac{1}{2}}\, |\, \sigma \rangle=0$. Thus, we see that the same expression vanishes identically. This allows us to express the correlator with the $L_{-\frac{3}{2}}L_{-\frac{3}{2}}$ mode inserted in terms of correlators with only integer-moded Virasoro generators inserted, which in turn can be evaluated by standard methods. Taking also the other modes into account, this leads to the following differential equation:
\begin{align}
%0 &=\langle\sigma\, |\, (L_{-1}^3\sigma)(u) \sigma(1)\,|\, \sigma \rangle-\frac{33}{16} \langle\sigma\, |\, (L_{-2}L_{-1}\sigma)(u) \sigma(1)\,|\, \sigma \rangle\nonumber\\
%&\qquad+\frac{93}{128} \langle\sigma\, |\, (L_{-\frac{3}{2}}^2\sigma)(u) \sigma(1)\,|\, \sigma \rangle-\frac{27}{64} \langle\sigma\, |\, (L_{-3}\sigma)(u) \sigma(1)\,|\, \sigma \rangle \\
&\left(\partial_u^3 -\frac{33(2u-1)}{16u(1-u)} \partial_u^2+\frac{3(5-192u+192u^2)}{256u^2(1-u)^2}\partial_u-\frac{15(2u-1)}{4096u^3(1-u)^3}\right)\langle\sigma\, |\, \sigma(u) \sigma(1)\,|\, \sigma \rangle = 0\ .
\end{align}

\subsection{Differential equation for the partition function}

Let us transform the obtained differential equation into a differential equation for the partition function of the theory by using \eqref{eq:2222=partition-fct}. We will exemplify this procedure again with the Yang-Lee and the Ising model. This yields
\be 
\left(\partial_u^2-\frac{2(2u-1)}{3u(1-u)}\partial_u+\frac{11(1-u+u^2)}{900u^2(1-u)^2}\right)Z(u)=0
\ee
and
\be 
\left(\partial_u^3-\frac{2(2u-1)}{u(1-u)}\partial_u^2+\frac{7-391u+391u^2}{192u^2(1-u)^2}\partial_u+\frac{23(2u-1)(u-2)(u+1)}{13824u^3(1-u)^3}\right)Z(u)=0
\ee
in the two cases, respectively.
We can then transform these differential equations into a modular differential equations in the variable $\tau$ by using the relation \eqref{eq:u tau relation}. For this, we use the Serre derivative, which is defined as
\be 
\mathscr{D}\equiv q \frac{\mathrm{d}}{\mathrm{d}q}-\frac{s}{12}E_2(\tau)=\frac{1}{2\pi i}\frac{\mathrm{d}}{\mathrm{d}\tau} -\frac{s}{12}E_2(\tau)\  ,
\ee
when acting on modular forms of weight $s$.
Here, we use the normalized Eisenstein series
\begin{align}
E_{2k}(\tau)&=\frac{1}{2\zeta(2k)}\sum_{(m,n) \ne (0,0)}\frac{1}{(m\tau+n)^{2k}}\ ,\qquad k \ge 2\ , \\
E_2(\tau)&=1+\frac{3}{\pi^2}\sum_{m \in \mathds{Z} \setminus \{0\}} \sum_{n \in \mathds{Z}}\frac{1}{(m\tau+n)^2}\ ,
\end{align}
where $\zeta(n)$ is the Riemann zeta function. Upon changing variables, the differential equations become
\begin{align}
0&=\left(\mathscr{D}^2-\frac{11}{3600} E_4(\tau)\right)Z(\tau)\ , \\
0&=\left(\mathscr{D}^3-\frac{107}{2304} E_4(\tau)\mathscr{D}+\frac{23}{55296}E_6(\tau)\right)Z(\tau)
\end{align}
for the Yang-Lee and Ising case, respectively. These coincide indeed with the well-known modular differential equations of the characters of these minimal models \cite{Eguchi:1986sb, Eguchi:1987qd, Mathur:1988rx, Gaberdiel:2008pr}.

\section{Discussion}\label{sec:discussion}
\subsection{Summary}
Let us summarise our findings. We developed a new method to compute correlation functions of twist fields in the symmetric product orbifold $\mathrm{Sym}^N(\mathcal{M})$. The method was applicable to correlators on the sphere whose covering space (defined by the monodromies of the fields around the twist fields) is also a sphere. In this case, we used the existence of a fractional Virasoro algebra acting on the twisted sector of the orbifold to derive differential equations for the correlators of twist fields. This determines the four-point functions of the theory up to an overall constant, which we fixed by imposing the correct factorisation behaviour into three-point functions. We generalised the method in several ways. We demonstrated that it can also effectively be applied to compute differential equations of correlation functions which have a higher-level null vector. We also showed that the analysis is essentially identical for higher-point functions. In all cases we have considered where the covering space is a sphere, the correlator takes the form
\begin{align}
\left \langle \prod_{i=1}^m\mathcal{O}_{g_i}(z_i) \right \rangle= \text{const.}\prod_{j=1}^m |a_j|^{-2h_j+\frac{c}{12}(w_j-1)}\prod_i |C_i|^{-\frac{c}{6}}\left \langle \prod_{i=1}^m\tilde{\mathcal{O}}(t_i) \right \rangle \ .\label{eq:precise relation base space covering space}
\end{align}
Here, $a_j$ and $C_i$ are coefficients of the covering map, see eqs.~\eqref{eq:Gamma expansions} and \eqref{eq:Gamma poles} (and all depend non-trivially on the coordinates $z_i$). The correlator on the right hand side is a correlator in the corresponding covering space, as determined by the group elements $g_1, \dots, g_m$. In all instances we have computed, this form is entirely fixed by the differential equations. 

We also showed that the method can still be applied if the covering space is a torus, provided that the seed theory of the symmetric product orbifold is a Virasoro minimal model. The resulting differential equation on four-point functions can be interpreted as a differential equation on the torus partition function of the theory and coincides with the modular differential equations of the theory.

\subsection[Relation to holography and the \texorpdfstring{$\mathrm{SL}(2,\mathds{R})$}{SL(2,R)} WZW model]{Relation to holography and the $\boldsymbol{\mathrm{SL}(2,\mathds{R})}$ WZW model}
Our results have interesting implications for the $\text{AdS}_3/\text{CFT}_2$ correspondence. The symmetric orbifold $\mathrm{Sym}^N(\mathbb{T}^4)$ was conjectured in \cite{Gaberdiel:2018rqv, Eberhardt:2018ouy, Eberhardt:2019ywk} to be dual to perturbative string theory on $\text{AdS}_3 \times \text{S}^3 \times \mathbb{T}^4$ with one unit of NS-NS flux, see also \cite{Argurio:2000tb}. The matching of the symmetry algebras in the correspondence was elucidated in \cite{Eberhardt:2019qcl} and in \cite{Dei:2019osr} it was found that null vectors in the CFT are BRST exact in the corresponding string worldsheet theory. The direct consequence of this is that the worldsheet correlators obey the same constraints as the CFT correlators. In view of our results, this hence allows to compute the $z$-dependence of correlators in the worldsheet theory. The twisted sector ground states are conjectured to be dual to vertex operators $V_{h_i}^{w_i}(z_i;t_i)$ in the $\mathrm{SL}(2,\mathds{R})_{k+2}$ WZW model, where $w_i$ becomes the spectral flow parameter on the worldsheet. For one unit of NS-NS flux, we have $k=1$.\footnote{For one unit of NS-NS flux, a better description of the worldsheet theory is given in terms of a $\mathrm{PSU}(1,1|2)_1$ WZW-model within the hybrid formalism \cite{Berkovits:1999im}. For more details, see \cite{Eberhardt:2018ouy}.} The vertex operators depend beyond the coordinate $x$ also on the worldsheet coordinate $z$.
It was recently argued in \cite{Eberhardt:2019ywk} that the worldsheet correlators on the sphere in the $\text{SL}(2,\mathds{R})_3$ WZW model with $\mathfrak{sl}(2,\mathds{R})$ spin $j=\tfrac{1}{2}$ have the following localised form\footnote{In the notation of \cite{Eberhardt:2019ywk}, we replaced $x_j$'s by $z_j$'s and $z_j$'s by $t_j$'s in order to be consistent with the conventions in the rest of this paper.} \footnote{By $\Gamma^{-1}(z_j)$, we mean the ramification point in the covering space that is mapped to $z_j$.} 
\begin{align}
\left \langle \prod_{j=1}^m V_{h_j}^{w_j}(z_j;t_j) \right \rangle=\sum_\Gamma W_\Gamma(t_4,\dots,t_m) \prod_{j=1}^m |a_j|^{-2h_j}\prod_{j=3}^{m-1}\delta^{(2)}\left(t_j-\Gamma^{-1}(z_j)\right) \ ,
\end{align}
where the sum runs over all covering maps with ramification indices $w_j$. The function $W_\Gamma$ depends only on the insertion points $t_4,\dots,t_m$ and the spectral flow numbers $w_1,\dots,w_m$. We have used the M\"obius symmetry in the dual CFT and on the worldsheet to fix $z_1=t_1=0$, $z_2=t_2=1$ and $z_m=t_m=\infty$, exactly as in the main text. This is the structure analogous to the one of correlators in the symmetric product orbifold, see the discussion in Section~\ref{subsec:large N}. Note also that the dependence on $h_j$ is fully fixed and has precisely the same form as in eq.~\eqref{eq:precise relation base space covering space}.
The string theory moduli space integral over $z_j$ trivialises thanks to the presence of the $\delta$-functions.

We can use the result of this paper to make a proposal for the unknown function $W_\Gamma(t_4,\dots,t_m)$. The requirement of the decoupling of BRST exact states (which is the analogue of the null-vector condition in the dual CFT, see \cite{Dei:2019osr}) suggests that the result has the form\footnote{Strictly speaking, \cite{Dei:2019osr} discussed that null vectors in the dual CFT correspond to BRST-exact states on the worldsheet only in a bosonic version of the correspondence. We have checked that the same conclusion is true in the supersymmetric incarnation of the duality for the level one descendants of the twisted sector ground states.}
\begin{align}
\left \langle \prod_{j=1}^m V_{h_j}^{w_j}(z_j;t_j) \right \rangle=\sum_\Gamma W_\Gamma\prod_{j=1}^m |a_j|^{-2h_j+\frac{1}{2}(w_j-1)}\prod_i |C_i|^{-1}\prod_{j=3}^{m-1}\delta^{(2)}\left(t_j-\Gamma^{-1}(z_j)\right) \ .
\end{align}
We have applied formula \eqref{eq:precise relation base space covering space} for $c=6$. Within the RNS formalism, there are also fermions that contribute to the correlator and the result we have given should already incorporate these fermionic contributions. Alternatively, one could work directly in the $\mathrm{PSU}(1,1|2)_1$ WZW model.
We have thus reduced the unknown part to only one constant $W_\Gamma$ per covering map.\footnote{We further suspect that $W_\Gamma$ is actually independent of $\Gamma$, since all covering maps are related by analytic continuation in the variables $t_i$ and $z_i$.} We expect that with this additional input, one can hope to solve the worldsheet theory on the sphere completely. It would be very interesting to explore this further in the future.
\subsection{Future directions}
Finally, we indicate possible future directions for this method to compute orbifold correlators.
\medskip

\paragraph{Higher genus.} We did not cover the higher genus case exhaustively. At least for rational CFTs or free theories like $\mathbb{T}^4$, treating the genus 1 case in full generality should not be out of reach. While one might not be able to constrain the full correlation functions through differential equations in general, one could conceivably compute the universal prefactor which arises through the conformal transformation to the covering space in the method of Lunin and Mathur \cite{Lunin:2000yv} and derive the analogue of \eqref{eq:precise relation base space covering space}. 
\paragraph{Supersymmetry and extremal correlators.} $\mathcal{N}=(4,4)$ supersymmetric theories like the symmetric orbifold of $\mathbb{T}^4$ have a protected subsector of extremal correlators. These are under much better control and their correlation functions are a lot simpler than those of the twisted sector ground states \cite{Lunin:2001pw, Dabholkar:2007ey, Gaberdiel:2007vu, Pakman:2009ab, Baggio:2012rr}. Thus one might hope to constrain them directly using differential equations. In the symmetric orbifold of $\mathbb{T}^4$, they also have the property that the sphere correlator is the \emph{only} connected contribution to the full correlator and thus they can be evaluated exactly.\footnote{This is almost true. There is one extremal correlator that receives a torus contribution. See \cite{Pakman:2009ab} for the precise statement.} There is also a further simplification: the covering map is a polynomial and can be written in closed form using hypergeometric functions \cite{Roumpedakis:2018tdb}. 
\paragraph{Other orbifolds.} An interesting extension of our result would be to extend the method to arbitrary orbifolds. It is certainly possible to consider also other permutation orbifolds of the form
\be 
\mathcal{M}^{\otimes N}/G\ ,
\ee
where the orbifold group acts by permutations. The twisted sectors of these orbifolds have the same form as for symmetric orbifolds and only the combinatorial factors change for the correlators $\langle \sigma_{[g_1]}(z_1) \cdots \sigma_{[g_2]}(z_2) \rangle$. It is much less obvious how to generalise the methods to arbitrary orbifolds, since the existence of the fractional Virasoro algebra is a special feature of permutation orbifolds. 
\section*{Acknowledgements}
We thank Juan Maldacena, Leonardo Rastelli, Martin Ro\v{c}ek, Konstantin Roumpedakis and Herman Verlinde for useful conversations. We thank Gleb Arutyunov and Sergey Frolov for helpful correspondence. We also thank Matthias Gaberdiel for suggesting the project to us, for useful comments on a preliminary version of the manuscript and for many enlightening discussions. The work of AD is supported by the Swiss National Science Foundation, and he acknowledges support by the NCCR SwissMAP, which is also funded by the Swiss National Science Foundation. LE gratefully acknowledges support from the Della Pietra family.

\appendix

\section{Identities for the covering map}
\label{app:Gamma-identities}
In this appendix, we derive various identities satisfied by the covering map. We take the covering map $\Gamma(z)$ to have the expansion
\be 
\Gamma(t)=z_j+a_j(t-t_j)^{w_j}+b_j(t-t_j)^{w_j+1}+c_j(t-t_j)^{w_j+2}+\mathcal{O}((t-t_j)^{w_j+2})
\ee
around the ramification points. Conventionally, we take $t_1=0$, $t_2=1$, $t_3=x$ and $t_4=\infty$, as well as $z_1=0$, $z_2=1$, $z_3=u$ and $z_4=\infty$, as in the main text. Implicitly, all the coefficients appearing in the expansion ($u$, $a_j$ and $b_j$) depend on $x$. We can also expand $\Gamma(z)$ around its poles $\ell_i$,
\be 
\Gamma(t)=\frac{C_i}{t-\ell_i}+\mathcal{O}(1)\ .
\ee
The number of poles is given by $n-w_4$, where $n$ is the degree of the map, which in turn is given by the Riemann-Hurwitz formula \eqref{eq:genus zero Riemann Hurwitz}.
\subsection{A relation between the coefficients}
In the main text, we need the relation
\be 
\sum_{j=1}^3 (w_j-1)\frac{\partial_x a_j}{a_j}-(w_4-1)\frac{\partial_x a_4}{a_4}- \frac{2b_3(w_3^2-1)}{a_3w_3}-2\sum_{i} \frac{\partial_x C_i}{C_i}=0  \label{eq:abC relation}
\ee
between the coefficients of the covering map. The fact that the fourth term appears with opposite sign in the series is because $t_4=\infty$, and our definition of $a_4$, see footnote~\ref{footnote:a4}. If we put $t_4$ at generic location, this minus sign would not appear.

Equation \eqref{eq:abC relation} is simply the statement that the sum of the residues of the function
\be 
f(t)=-\partial_t\left(\frac{\partial_t^2\Gamma(t)}{(\partial_t\Gamma(t))^2}\right)\partial_x \Gamma(t)
\ee
equals zero. The function only has poles at either the insertion points or the poles of $\Gamma(t)$. The residues are
\begin{subequations}
\begin{align}
\mathop{\text{Res}}_{t=t_j} f(t)&=(w_j-1)\frac{\partial_x a_j}{a_j}\ , \qquad j=1,\, 2\ , \\
\mathop{\text{Res}}_{t=t_3} f(t)&=(w_3-1)\frac{\partial_x a_3}{a_3}-\frac{2b_3(w_3^2-1)}{a_3w_3}\ ,  \\
\mathop{\text{Res}}_{t=t_4} f(t)&=-(w_4-1)\frac{\partial_x a_4}{a_4}\ , \\
\mathop{\text{Res}}_{t=\ell_{i}}f(t)&=-2\sum_{i} \frac{\partial_x C_i}{C_i}\ .
\end{align}
\end{subequations}
Hence \eqref{eq:abC relation} follows. 

In Section~\ref{sec:higher-point-and-multi-cycle-corr} we will need a slight generalization of eq.~\eqref{eq:abC relation}. If there are $m$ insertion points of the correlator, we find 
\be 
\sum_{j=1}^{m-1} (w_j-1)\frac{\partial_{x_a} a_j}{a_j}-(w_{m}-1)\frac{\partial_{x_a} a_{m}}{a_{m}}- \frac{2b_{a+2}(w_{a+2}^2-1)}{a_{a+2} w_{a+2}}-2\sum_{i} \frac{\partial_{x_a} C_i}{C_i}=0\ , \label{eq:abC relation-many}
\ee
where $x_1, \dots, x_{m-3}$ are the cross-ratios in the covering space and for $a=1, \dots, m-3$, we have $\Gamma(x_a) = u_a$, with $u_a$ labeling the cross-ratios in the base space and $\Gamma(t)$ being defined by a slight generalization of \eqref{eq:Gamma expansions five}. Here we put in analogy to the four-point function case $x_1=t_1=0$, $x_2=t_2=1$ and $x_m=t_m=\infty$.  

\subsection{A second relation between the coefficients}

In Section~\ref{sec:more-differential-equations}, in order to derive the second order equation, we need a further relation between the coefficients.
To derive it, we consider the function
\be 
f(t)=  \left(2\frac{\partial_t^3\Gamma}{\partial_t\Gamma}-3\frac{(\partial_t^2 \Gamma)^2}{(\partial_t\Gamma)^2}\right) \frac{t (t-1)}{t-x} \ .
\ee
Note that the term in parenthesis is the Schwarzian derivative.
This function has again only poles at the insertion points and at the poles of $\Gamma(t)$. The residues are
\begin{subequations}
\begin{align}
\mathop{\text{Res}}_{t=0} f(t)&= \frac{1-w_1^2}{x} \ , \\
\mathop{\text{Res}}_{t=1} f(t)&= \frac{1-w_2^2}{1-x}  \ , \\
\mathop{\text{Res}}_{t=x} f(t)&= \frac{1}{a_3^2 w_3^2} \Bigl[a_3^2 w_3^2 \left(1-w_3^2\right)  -4 a_3 \hspace{1pt} c_3 \hspace{1pt} w_3 \left(w_3^2-4\right) (x-1) x \nonumber \\
& \hspace{45pt} -2 a_3 \hspace{1pt} b_3 \hspace{1pt} w_3 \left(w_3^2-1\right) (2 x-1) + b_3^2 (w_3+1)^2 (2 w_3-5) (x-1) \hspace{1pt} x \Bigr] \ , \\
\mathop{\text{Res}}_{t=\infty} f(t)&= w_4^2-1 \ , \\
\mathop{\text{Res}}_{t=\ell_i} f(t)&= 0 \ .
\end{align}
\end{subequations}
Using that the sum of all residues vanishes, we can express $c_3$ in terms of the other constants (note that on top of this, we can also solve \eqref{eq:abC relation} for $b_3$ and insert it), 
\begin{multline}
c_3 =  -\frac{a_3 w_3}{4 \left(w_3^2-4\right) (x-1) x} \Biggl(-\frac{b_3^2 (w_3+1)^2 (2 w_3-5) (x-1) x}{a_3^2 w_3^2}\\
 +\frac{2 b_3 \left(w_3^2-1\right) (2 x-1)}{a_3 w_3}  +\frac{w_1^2-1}{x}+\frac{w_2^2-1}{1-x}+w_3^2-w_4^2 \Biggr) \ .\label{eq:abC relation-bis} 
\end{multline}

\bibliographystyle{JHEP}
\bibliography{bib}
\end{document}